\documentclass{aa} 
\usepackage{graphicx}
\usepackage{natbib}

\usepackage{txfonts}

\newcommand{\msun}{\mbox{$\rm M_\odot\,$}}

\newcommand{\lsun}{\mbox{$\rm L_\odot\,$}}

\newcommand{\mum}{\mbox{$\rm \mu m\,$}}
\def\herschel{$\it{Herschel}$\,}
\def\deg{\hbox{$^\circ$}}
\def\arcmin{\hbox{$^\prime$}}
\def\arcsec{\hbox{$^{\prime\prime}$}}

\begin{document}

\title{Unifying low- and high-mass star formation through density-amplified hubs of filaments} \titlerunning{Young clusters,
  high mass stars and hubs} \subtitle{The highest mass stars
  ($>$100\msun) form only in hubs}

   \author{M. S. N. Kumar\inst{1}, P. Palmeirim\inst{1}, D. Arzoumanian\inst{1},
          \and
          S. I. Inutsuka\inst{2} 
          }

   \institute{Instituto de Astrof\'{i}sica e Ci\^{e}ncias do Espaço,
     Universidade do Porto, CAUP, Rua das Estrelas, PT4150-762 Porto,
     Portugal \and Department of Physics, Nagoya University, Furo-cho,
     Chikusa-ku, Nagoya, Aichi 464-8602, Japan
     \\ \email{nanda@astro.up.pt} }

   \date{Received April 23, 2020; accepted July 31, 2020}

 
  \abstract { Star formation takes place in giant molecular clouds,
    resulting in mass-segregated young stellar clusters composed of
    Sun-like stars, brown dwarfs, and massive O-type(50-100\msun)
    stars.}{We aim to identify candidate hub-filament systems (HFSs)
    in the Milky Way and examine their role in the formation of the
    highest mass stars and star clusters.}{The \herschel survey HiGAL
    has catalogued about 10$^5$ clumps. Of these, approximately 35000
    targets are detected at the 3$\sigma$ level in a minimum of four
    bands. Using the DisPerSE algorithm we detect filamentary
    skeletons on 10$\arcmin \times 10\arcmin$ cut-outs of the SPIRE
    250\mum images (18\arcsec\, beam width) of the targets. Any
    filament with a total length of at least 55\arcsec (3 $\times$
    18\arcsec) and at least 18\arcsec\, inside the clump was
    considered to form a junction at the clump. A hub is defined as a
    junction of three or more filaments. Column density maps were
    masked by the filament skeletons and averaged for HFS and non-HFS
    samples to compute the radial profile along the filaments into the
    clumps.}{Approximately 3700~(11\%) are candidate HFSs, of which
    about 2150~(60\%) are pre-stellar and 1400~(40\%) are
    proto-stellar. The filaments constituting the HFSs have a mean
    length of $\sim$10-20\,pc, a mass of $\sim$5$\times$10$^4$ \msun ,
    and line masses ($M/L$) of $\sim$2$\times$10$^3$ \msun
    pc$^{-1}$. All clumps with L$>$$10^4$\lsun and L$>$$10^5$\lsun at
    distances within 2kpc and 5kpc respectively are located in the
    hubs of HFSs. The column densities of hubs are found to be
    enhanced by a factor of approximately two (pre-stellar sources) up
    to about ten (proto-stellar sources).}{All high-mass stars
    preferentially form in the density-enhanced hubs of HFSs. This
    amplification can drive the observed longitudinal flows along
    filaments providing further mass accretion. Radiation pressure and
    feedback can escape into the inter-filamentary voids. We propose a
    `filaments to clusters' unified paradigm for star formation, with
    the following salient features: (a) low-intermediate-mass stars
    form slowly (10$^6$yr) in the filaments and massive stars form
    quickly (10$^5$yr) in the hub, (b) the initial mass function is
    the sum of stars continuously created in the HFS with all massive
    stars formed in the hub, (c) feedback dissipation and mass
    segregation arise naturally due to HFS properties, and explain the
    (d) age spreads within bound clusters and the formation of
    isolated OB associations.}


   \maketitle
%

\section{Introduction}

Star formation in giant molecular clouds produces mass-segregated
clusters, with the most massive stars located at the centre
\citep{LadaLada2003,Zwart2010}.  The mass function of the resulting
stars is similar to the Salpeter mass function. Once formed, massive
stars are thought to drive significant feedback and produce ionised
(HII) regions \citep{Deharveng2010,Samal2018} eventually clearing the
natal molecular cloud in $\sim$3-5\,Myr \citep{LadaLada2003}. Typical
formation timescales for high- and low-mass stars are $\sim$10$^5$yr
\citep{BehrendMaeder2001} and 10$^6$yr respectively; if massive stars
form first, the feedback can inhibit the formation of the low-mass
stars or even halt it by blowing away the natal cloud. If low-mass
stars form prior to high-mass stars \citep{Kumar2006}, this effect can
be negated and some properties of nearby star forming regions such as
the Orion or the Carina nebulae can be explained.

Not all star formation results in dense clusters with OB stars, and
not all clusters remain bound following gas dispersal
\citep{LadaLada2003,Zwart2010,Krumholz2019}. Star formation in nearby
regions such as Taurus, Perseus, Chameleon, and Ophiucus lack the
O-stars that can be seen in regions such as Orion, Rosette, M8, W40,
and Carina. An intriguing observational property of dense clusters
such as the Orion Nebula Cluster (ONC) is the finding by
\citet{Palla2007} that old stars are found in the midst of young
clusters. The ONC is generally considered to have an age of
$\sim$1\,Myr \citep{LadaLada2003}. However, evidence of an extended
star formation history (1.5--3.5\,Myr) displaying some dependence on
the spatial distribution of stars has also been found
\citep{Reggiani2011}.

While lower mass star formation that may lead to dispersed populations
in the Milky-Way is reasonably well understood
\citep{McKee2007,Kennicutt2012}, the challenge so far is in arriving
at a universal scenario of star formation that can reconcile the
observational properties of a well-studied cluster such as the Carina
Nebula \citep{Smith2006}, explaining (a) the formation of the highest
mass stars such as Eta Carinae ($\sim$120\,\msun), (b) mass
segregation, and (c) the origin of the full spectrum of initial mass
function, especially accounting for the difference in formation times
between high- and low-mass stars, and the feedback effects.

The most massive stars catalogued in the Milky Way are typically
100-150\msun, though some authors have claimed the existence of
200-300\msun stars in the R136 cluster
\citep{Crowther2010}. Theoretically, even though radiation pressure
was thought to set an upper limit on the formation of the most massive
star, it is now argued that such a limit does not exist
\citep{Krumholz2015}. The proposition that high-mass stars form as
scaled-up versions of low-mass stars stems from certain observational
similarities between them, such as outflows
\citep{Shepherd1996}. Nevertheless, the scaled-up idea is largely
propagated by theoretical models based on turbulent core accretion
\citep{McKeeTan2003}, proposing mechanisms to dispel radiation
pressure, and even utilising feedback to suppress fragmentation
\citep{Krumholz2006}. There remains a persistent lack of evidence to
explain the formation of the average O-stars (30-50\msun) with main
sequence lifetimes of a few million years, and even more, giving rise
to stars of $>$100\msun.  Furthermore, how the necessary mass
reservoir is assembled and tens of solar masses are accreted (rate
$\dot{M} \sim 10^{-4}-10^{-2}$\,\msun~yr$^{-1}$) on a timescale that
is widely believed to be about 10$^5$\,yr is also largely unclear
\citep{Kennicutt2012,BehrendMaeder2001,Krumholz2015}.  A handful of
claims of discs in targets representative of $\sim$20\msun stars are
the best observational evidence of the proto-stellar stage
\citep{Zapata2019}, and observations searching for massive pre-stellar
cores have declared them to be the `holy grail'
\citep{Motte2018}. Observationally, there remains no evidence of disc
accretion at a rate of $\sim$10$^{-3}$\msun\,yr$^{-1}$, and the topic
is an ongoing challenge.

The idea that the interstellar medium (ISM) of the Milky Way is
organised in filamentary structures and bubbles \citep{Inutsuka2015}
was proposed decades ago \citep{Heiles1979,Schneider1979}.
Following the Herschel space mission, this view has matured, and the
properties of filamentary structures have been quantified. It is now
believed that the cold ISM, held as the birthplace of stars, is mostly
found to be organised in filamentary structures
\citep{Andre2010}. More than 80\% of the dense gas mass (above a
column density representing A$_v>$7) in the nearby star forming
regions is shown to be in the form of filaments
\citep{Konyves2015,Arzoumanian2019,Konyves2019}. The star forming
clouds in much of the Milky Way are now viewed to be filamentary in
nature \citep{Inutsuka2015}. Dense filamentary structures in the Milky
Way disc have been uncovered using Galactic-plane surveys such as
ATLASGAL and Hi-GAL \citep{Li2016,Mattern2018,Schisano2020}, with a
wide range in lengths (a few to 100\,pc) and line masses (a few
hundreds to thousands of \msun pc$^{-1}$).

In a molecular cloud, these filamentary structures inevitably overlap
to form a web, creating junctions of filaments. \citet{Myers2009}
identified such junctions for the first time, defining them as hubs,
objects of low aspect ratio and high-column density, in contrast to
filaments that have high aspect-ratio and lower column densities. This
author also showed that the nearest young stellar groups are
associated with hubs that radiate multiple filaments and pointed out
that such a pattern was also found in infrared-dark clouds. Directed
by this observational association of hubs and clusters, many authors
have studied hub-filament systems (HFSs) as possible progenitors of
proto-clusters and high-mass star formation \citep{Schneider2012,
  Mallick2013}. Young stellar clusters (YSCs) can be split up
according to mass, with the highest mass stars located at the centre,
which prompted other authors to investigate high-mass star formation
in HFSs \citep{Liu2012, Peretto2013}. These studies clearly
demonstrate the role of HFSs as important observational targets to
understand both the formation of high-mass stars and YSCs.
Observations of HFSs have uncovered longitudinal flows
\citep{Peretto2013,Peretto2014,Williams2018} within filaments with
flow rates of $\sim$10$^{-4}$--10$^{-3}$\msun yr$^{-1}$
\citep{Chen2019,TrevinoMorales2019}. Because such flows are found to
converge to a cluster of stars \citep{Chen2019,TrevinoMorales2019}, it
has been argued that the flow, triggered by a hierarchical global
collapse, provides sufficient flow rates to form massive stars. Hubs
with either a few \citep{Williams2018,Chen2019} or a large network of
filaments \citep{TrevinoMorales2019} both report similar flow rates,
which leads us to pose the following questions; what is the difference
between such HFSs with few filaments and large network of filaments?
Are the flows driven by a pre-existing massive star / clusters
gravitational potential drives or if the formation of the massive star
is a consequence of the flows? These observations of early to
intermediate stages of cluster formation will always be plagued by the
uncertainty of whether or not fragmentation-induced starvation limits
the formation of the most massive stars \citep{Peters2010}.

Here we approach the problem in reverse order by asking whether or not
the highest luminosity (therefore, highest mass) stars that have
recently formed are associated with HFSs, and if so, what it is that
makes these regions unique. The \herschel Hi-GAL survey has led to an
unprecedented and unbiased sample of star-forming clumps in the entire
Milky Way disc.  The low aspect ratio and high column density of hubs
can make them appear similar to any star-forming clump of dense
gas. However, not all clumps can be hubs; especially when considering
the results from \citet{Myers2009}, that hubs coincide with the
centres of stellar groups. The fraction of cluster-forming hubs must
be quite small when compared to all the clumps in a giant molecular
cloud. Additionally, for an observer, line-of-sight coincidences of
filamentary structures may mimic a hub-like structure. In the
filamentary paradigm of molecular clouds, there are main filaments,
sub-filaments, and striations, as well as junctions of each of these
structures which can in principle also be called a hub. By this
definition, the hubs defined by \citet{Myers2009} represent junctions
of main filaments, and there should be a hierarchical distribution of
hubs depending on the type- and number of elements intersecting to
form a hub. The nature of the star formation that takes places in hubs
will then depend upon the nature of these junctions, resulting from
the density and number of intersecting filamentary structures.

Therefore, it is unclear whether or not every HFS leads to the
formation of high-mass stars and/or clusters and the current study is
just the beginning of what we may learn by observing these
systems. However, given the unambiguous importance of HFSs, it is
necessary to first identify such systems in an unbiased way. This work
aims to identify candidate HFSs towards the inner Galactic plane using
Hi-GAL data. The analysis is different from other `filament
catalogues' of the Galactic plane \citep[e.g.][]{Li2016,Schisano2020}
because it does not search for filaments but is looking at filaments
merging into clumps. Section 2 describes the data sets used to conduct
the search. Section 3 details the definitions and methods of
identification of filamentary structures and hubs and Sect. 4 reports
the results and describes the output products. Based on the results,
we present and discuss a `filaments to clusters' paradigm for star
formation in Sect. 5. In Sect. 6 we compare this paradigm with the
literature and some archival data of two nearby regions of cluster
formation namely NGC2264 and W40. In Sect.\,7, we compare the HFS
paradigm with other models of cluster formation. In Sect.\,8 we
discuss the implications of the HFS paradigm for the hierarchy of HFSs
and observations of massive discs and pre-stellar cores, and its
relevance to the formation of very massive stars, feedback, and
triggered star formation.

\section{Observational data}

 We used 250$\mum$ maps, column density maps, and the clumps catalogue
 from the Herschel Hi-GAL survey to search for filamentary structures
 and hubs.  The Herschel Infra-red Galactic plane survey
 \citep[Hi-GAL;][]{Molinari2010,Molinari2016} covered the inner part
 of the Galactic Plane ($68^{\circ} \ge l \ge -70^{\circ}$ and $|b|
 \le 1^{\circ}$) using PACS \citep{Poglitsch2010} 70~$\mu$m and
 160~$\mu$m and SPIRE \citep{Griffin2010} at 250~$\mu$m, 350~$\mu$m,
 and 500~$\mu$m simultaneous imaging in all five bands. These data
 were reduced using the ROMAGAL data-processing code, for both PACS
 and SPIRE (see Traficante et al. 2011 for
 details)\nocite{Traficante2011}. Images of the four bands from
 160~$\mu$m to 500~$\mu$m were then used to compute the column density
 and the dust temperature ($T_\mathrm{d}$) maps. A catalogue of Hi-GAL
 clumps was prepared by \citet{Elia2017}, which contain $\sim$10$^5$
 sources. This catalogue lists various properties of the detected
 clumps such as the full width half maximum, luminosity, distance,
 surface density, and proto or pre-stellar nature of the
 clump. Identification of clumps is based on the photometric
 catalogue, the details of which can be found in
 \citet{Molinari2016}. The sources are extracted using bi-dimensional
 Gaussian fits to the source profile using the CuTEx algorithm
 \citep{Molinari2011}; we strived to achieve completeness in each
 band. Typically, a 90\% completeness limit on the Hi-GAL clump
 catalogue yields 5\msun clumps at 1\,kpc distance.
 
In this work, the distances provided by the clump catalogue
\citep{Elia2017} have been updated based on the distances provided by
\citet{Urquhart2018}. For those clumps where distances are not
available, if they are enclosed in the star-forming region of a known
distance, that distance is assumed.

\subsection{Targets and data selection}
In order to identify HFSs, it is necessary to examine the filamentary
structures around every known Hi-GAL clump, and evaluate if it
represents an intersection of filamentary structures. Given that HFSs
are known to be associated with YSCs, and our interest in studying
high-mass star and proto-cluster formation, the requirement here is in
the robustness of the target clumps rather than completeness. To
obtain a robust sample of clump targets, we conducted a quality
analysis and cut to the sources in the original Hi-GAL catalogue. The
quality-controlled sample-selection criteria are as follows:
\begin{itemize}

\item Detection at all four bands from 160$\mum$ to 500$\mum$ (71\% of the
  total sample: 44,686 sources).
\item Peak position precision between two consecutive bands within 3$\sigma$
  of the positional offsets of all sources (of any flux) in a given
  band (60\% of the sources remaining - 37,494).
\item  3$\sigma$ flux detection in all four bands (55\% - 34,575).
\end{itemize}

The total number of target clumps satisfying all the above criteria
is therefore 34,575. There are 145 sources that are saturated in one
or all bands; we modelled and corrected these, and have
included them in the above quality-controlled sample.

Given that the targets are located at a wide range of distances from
within the solar neighbourhood all the way up to 10-12\,kpc, angular
resolution is an important criterion to maximise the output of our
filamentary-feature detection. Therefore, we perform our HFS search
using the 250$\mum$ images that are considered to be the best
representation of column density while having a superior angular
resolution (18.2\arcsec\, beam fwhm) compared to the column density
maps at 36\arcsec\, resolution.  We extracted 10\arcmin $\times$
10\arcmin\, images of the calibrated 250\mum maps around each of the
34,575 targets to identify the filamentary structures.

\section{Identifying filamentary structures and hubs}

\subsection{The DisPerSE algorithm} 

The Discrete Persistent Structures Extractor (DisPerSE) software
\citep{Sousbie2011a} was used to identify filamentary skeletons in
each of the above image cut-outs.  DisPerSE is designed to identify
persistent topological features such as filamentary structures, voids,
and peaks. It works by connecting critical points such as maxima and
minima with integral lines along the gradients in a given
map. Critical points correspond to a zero gradient on the map. One of
the important inputs for the algorithm to run is `persistence level'
which is the absolute difference between the values of the critical
points, or in other words, between the maxima and minima along a
gradient that is connected by a single integral line called an `arc'.

\begin{figure}
   \includegraphics[width=0.5\textwidth]{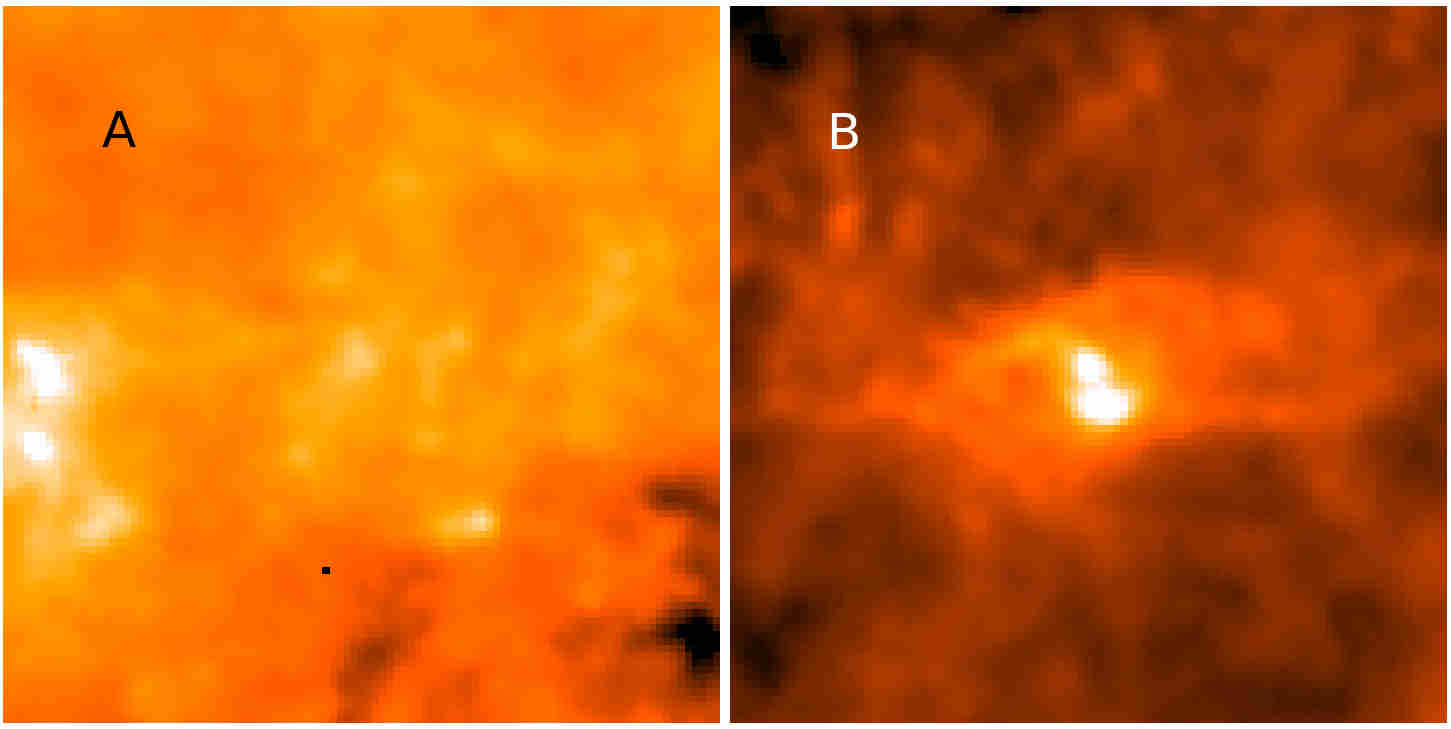}
   \includegraphics[width=0.5\textwidth]{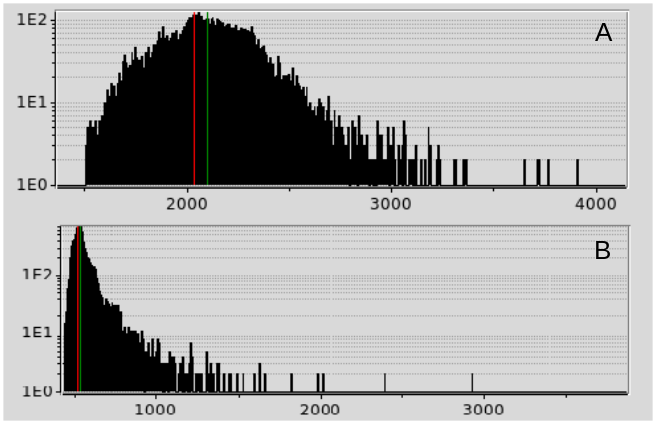}
   \caption{Pixel value distribution histogram examples of
     10\arcmin$\times$10\arcmin \ cut-out maps of (A) Background-dominated
     image, and (B) source-clump-dominated image. The x-axes of the
     histograms denote the pixel intensity values in MJy/Sr. The red
     and green vertical lines mark the mode and midpoint
     respectively.}
              \label{}%
    \end{figure}
    
 \begin{figure}
   \includegraphics[width=0.5\textwidth]{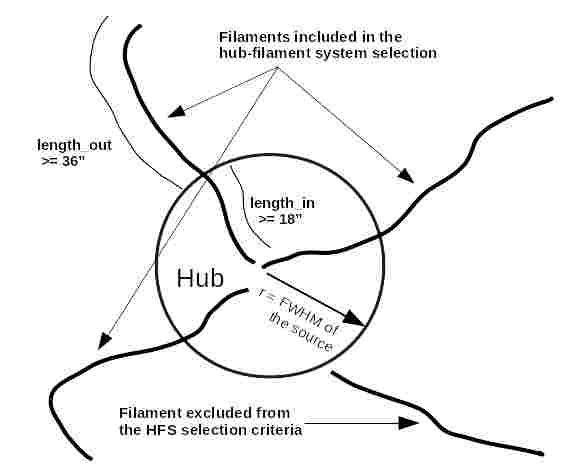}
   \caption{Hub-filament systems selection criteria }
              \label{}%
    \end{figure}

 \begin{figure*}
   \centering
   \includegraphics[width=0.9\textwidth]{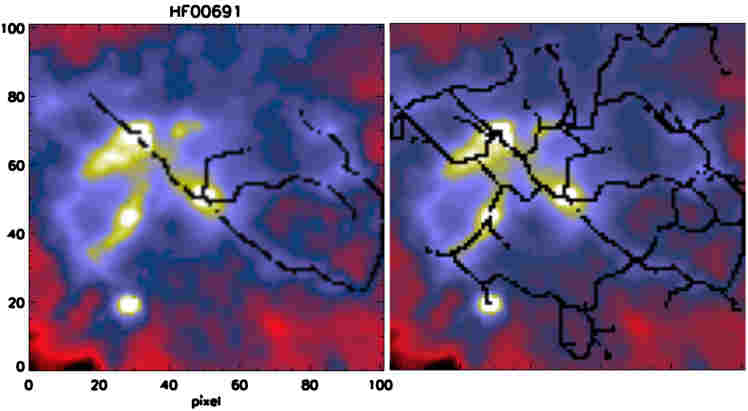}
   \caption{Example of an HFS candidate: skeletons passing the HFS
     criteria (left panel) and all skeletons selected by DisPerSE
     (right panel) are overlaid on a 250\mum image.}
              \label{}%
    \end{figure*}

 \begin{figure*}
   \centering
   \includegraphics[width=0.9\textwidth]{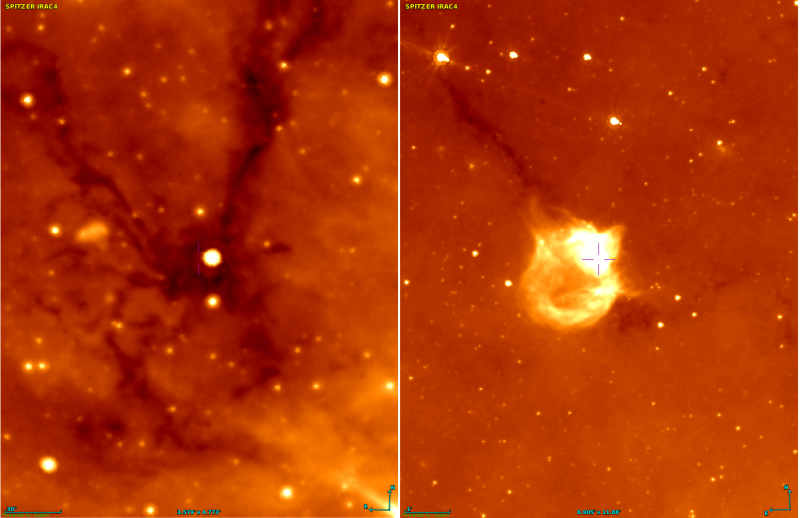}
   \caption{Hub-filament examples as viewed by the {\em Spitzer} 8\mum
     images: Candidate HFSs may be composed of filaments that are
     infrared dark (left) or bright, or a combination of infrared-dark
     and bright features (right). }
              \label{}%
    \end{figure*}

\vskip 2mm
\noindent {\em Persistence Level and Background:} The target clumps
are spread along the Milky Way plane, so the background varies
significantly between clumps. The mean intensity value of the
background (mode of the intensity, explained further ahead) on the
250\mum image (10\arcmin$\times$10\arcmin cut-outs) varies between 10
and $\sim$18000, with a standard deviation of $\sim$800, all in units
of MJy/Sr. This corresponds to column density variations between
9$\times$10$^{19}$cm$^{-2}$ and 1.5$\times$10$^{23}$cm$^{-2}$ on the
cut-outs. These numbers demonstrate that in some cases the cut-out
image is filled mostly by the emission from the dense clump and in
some cases by the emission surrounding the target clump. Additionally,
depending on the intensity of the source, the background-to-source
ratio plays an important role in detecting significant filamentary
features. We experimented in running DisPerSE on 250$\mum$ images and
an unsharp masked version where the background was subtracted by a
smoothing equal to 1.5 times the beam width. Comparing the results, it
was found that DisPerSE can efficiently detect features even in a
noisy image without background subtraction, provided a correct
background is determined. In Fig.1 we display the pixel distribution
histograms of two images, demonstrating the nature of pixel
distribution in the image that is dominated by background emission and
another where the dense clump is bright compared to the background. By
definition, mode is the peak of the pixel distribution histogram,
which represents the `background' or `most common' value, shown by a
red line. Another useful quantity is the `midpoint' which is estimated
by integrating the histogram and computing by interpolation of the
data value at which exactly half the pixels are below that data value
and half are above it. In IRAF, both mode and midpoint are computed in
two passes, unlike statistics such as the mean, standard deviation,
and so on. In the first pass the standard deviation of the pixels is
computed and used with the bin-width parameter to compute the
resolution of the data histogram. In the second pass, the mode is
computed by locating the maxima of the data histogram and fitting the
peak by parabolic interpolation. The midpoint is typically larger than
the mode, as shown by the red and green lines in Fig.1.

We define the persistence level for each source as five times the
standard deviation of the pixels below the midpoint value. We note
that the midpoint is fairly close to the mode or the background, while
ensuring that half the pixels in the distribution histogram are
considered. This is a conservative choice, made after experimenting by
setting the persistence level as the standard deviation of the pixels
below the mode and midpoint. All detected skeletons are therefore
above the 5$\sigma$ level of the background variations in each
target. Many of the Hi-GAL clumps are located in the midst of giant
molecular clouds or HII regions, where the estimated background
represents the ambient column density of the cloud or nebulosity. In
that case the chosen persistence level is set to pick up skeletons
well above the inter-cloud column density fluctuations.

 The DisPerSE algorithm is executed on the 250\mum cut-out image of
 each target using the respective persistence level. This is
 accomplished by running the main task {\em mse} within the
 algorithm. In the next step, DisPerSE builds `skeletons' of crests
 traced by the arcs above the given persistence level. Aligned arcs
 are assembled into individual skeletons, with the alignment defined
 by a critical angle between two consecutive arcs. We set this angle
 at 55\deg\,, similar to previous methods of filament detection
 (e.g. Arzoumanian+2019). Each skeleton assembled is tagged by a
 number that is based on the order in which it was detected: the
 skeleton picked up first by the algorithm will be assigned the value
 1, and so on. This skeleton tag is a useful indicator of the
 prominence of the detected features. This step is accomplished by
 running the task {\em skelconv}. The result of the above two steps is
 a {\em fits} file with all identified skeletons above a certain
 persistence level, tagged with respective numbers.

\subsection{Selecting the hub-filament systems}

The working definition of a filament is that it should have an aspect
ratio of at least three, $L_{fil} / W_{fil} \ge 3,$ as defined by previous
exercises \citep{Arzoumanian2019,Konyves2019}. Given that the width of
the filament $W_{fil}$ is unresolved in almost all the cases for the very
distant targets in this study, the width is defined by the beam size
at 250$\mum$(18.2\arcsec); therefore, a skeleton should have a length
$L_{fil} \ge 55\arcsec$ to be considered as a filament. The definition
of a HFS is sketched in Fig.\,2. At the outset, a hub
is defined as a junction of three or more skeletons at the source
centre. The source size varies as the full width at half maximum (FWHM)
of the detected clump. Any filament is considered to meet at the
source if at least one beam width, represented by three native pixels
(6\arcsec$\times$3=18\arcsec) of the skeleton, falls within the source
FWHM. { We note that DisPerSE is highly effective in picking up
  every filamentary structure, not all of which can be considered
  prominent by a visual inspection, especially when dealing with weak
  clumps or targets located at large distances.} Also, sometimes, the
algorithm selects linear structures at the edges of the images that
should be clipped away. Therefore, we imposed two additional criteria;
(a) that every pixel of the skeleton should have a minimum intensity of
3$\sigma$ ($\sigma$ of pixels below the midpoint value, see Sect.3.1)
in order to be included in the HFS criteria, and (b) only the first
half of detected skeletons (identified by the number tagged by
DisPerSE to the skeleton) are considered. The second constraint is
effective in clipping away unwanted edge-of-the-image structures and
very small skeleton branches of low intensity. All targets that
satisfied the above two criteria along with that defined in Fig.\,2.
were selected as candidate HFSs.

The above criterion was not applied to saturated sources because the
intensities in the central regions are modelled by Gaussian fitting
and the saturated sources generally fill a large fraction of the
10\arcmin$\times$10\arcmin\, cut-outs, strongly influencing the
background value by the large-scale nebular emission. Instead, the
only constraint imposed was that a filament should have non-zero
length within the FWHM of the clump in order to be considered as a
hub. The FWHMs of the saturated sources are also larger than those of
the non-saturated clumps. Visual examination shows dense networks of
filaments in every source (see Fig.\,13).

A catalogue was assembled listing the indicative properties of the
filaments and the associated clump properties. The total filament
lengths were separated into length inside and outside the clump. An
indicative column density of the respective parts of the skeleton was
computed by reading out the values of each skeleton pixel from the
Hi-GAL column density maps. Also, the total column density within a
circle of radius equal to the source FWHM was computed. For each
target, the mode from the column density maps of 10\arcmin
$\times$10\arcmin (similar to the 250\mum\, maps) was used as a
background value that was subtracted from the column densities read
out for each skeleton pixel.  A sample of the HFS catalogue is shown
in Table 1 and the full list is made available via CDS and Vizier
platforms.

\begin{table*}
\caption{\label{t1}Catalogue of hub-filament system candidates}
\centering
\begin{tabular}{lccccccccc}
\hline\hline
HFS & l & b & dist & n$_{skel}$ &  NH$_{2}^{out}$ & len$_{out}$ & M$^{FIL}$ & Line Mass \\
    & deg & deg & kpc & & 10$^{-23}$cm$^{-2}$ & pc & 10$^3$\msun & 10$^3$\msun\,pc$^{-1}$ \\
\hline
  HF00012  & 9.959827  &  -0.208277 &  3.5 &   3    &    34.1  &   11.7  & 75.3  &   6.4 \\
  HF00027  & 9.90676   &  0.386485  &  3.0 &   3    &    9.1  &   4.7   & 6.9   &  1.5 \\
  HF00032  & 9.89602   &  -0.418992 &  3.5 &   4    &    41.0   &   16.0  & 124.4 &     7.7 \\
  HF00035  & 9.875734  &  -0.74992  &  3.1 &   3    &    42.3  &   13.7  & 96.8  &   7.1 \\
  HF00036  & 9.873115  &  -0.748085 &  3.1 &   3    &    49.9  &   14.4  & 120.2 &   8.3 \\
  HF00038  & 9.870444  &  0.898442  &  3.0 &   4    &    18.6  &   13.7  & 41.5  &   3.0 \\
\hline
\end{tabular}
\tablefoot{The catalogue columns are: 1) HFS: running index, 2)
  Galactic longitude in degrees, 3) Galactic latitude in degrees, 4)
  distance in pc, 5) n$_{skel}$ : number of skeletons at the junction,
  6) NH$_{2}^{out}$: background (defined as the mode of the
  10\arcmin$\times$10\arcmin image) subtracted column density of the
  skeleton lying outside the source FWHM, 7) len$_{out}$: total length
  of the skeletons outside the source FWHM, 8) M$^{FIL}$: Mass of the
  filament computed using the NH$_{2}^{out}$. This is to avoid
  contribution from the clump, and 9) Line Mass in M$\odot$/pc : Line
  mass of the filament defined as M$^{FIL}$/len$_{out}$.}
\end{table*}

The filament mass was computed using the measured column density
NH$_{2}^{out}$ and using the formula M$^{FIL}$ = NH$_{2}^{out}$
$\times$ area$_{Fil}$ $\times\, \mu_{H_2} \times\, m_H $ where
$\mu_{H_2}$ is the mean molecular weight per H$_2$ taken as 2.33 and
m$_H$ is the atomic weight of hydrogen. The filament length is
multiplied by one pixel width (6\arcsec) to obtain the area$_{Fil}$.
Additionally, we also provide overlays of HFS skeletons on the 250\mum
images for visualisation of the candidate systems as shown in Fig. 3a.

\subsection{Column density profile of hub-filament systems}

The DisPerSE output skeletons were used as a mask on the corresponding
column density map cut-outs of each target, allowing us to measure it
along the filaments and clumps located on the filaments. The result is
shown in the left panel of Fig. 5. These masked column density maps of
all HFS and non-HFS targets were averaged to produce the middle and
right panel images in Fig. 5. In this averaged image, the central
pixel corresponds to the clump centre. Next, we computed the average
column density value in concentric circles of one pixel width with
respect to this centre, to produce circularly averaged radial profiles
shown in Fig.\,11. The standard deviation of this circular average is
used as the corresponding error in the figure. Such profiles were
computed for pre-stellar and proto-stellar sources separately in the
HFS and non-HFS groups.

\begin{figure*}
   \centering
   \includegraphics[width=0.9\textwidth]{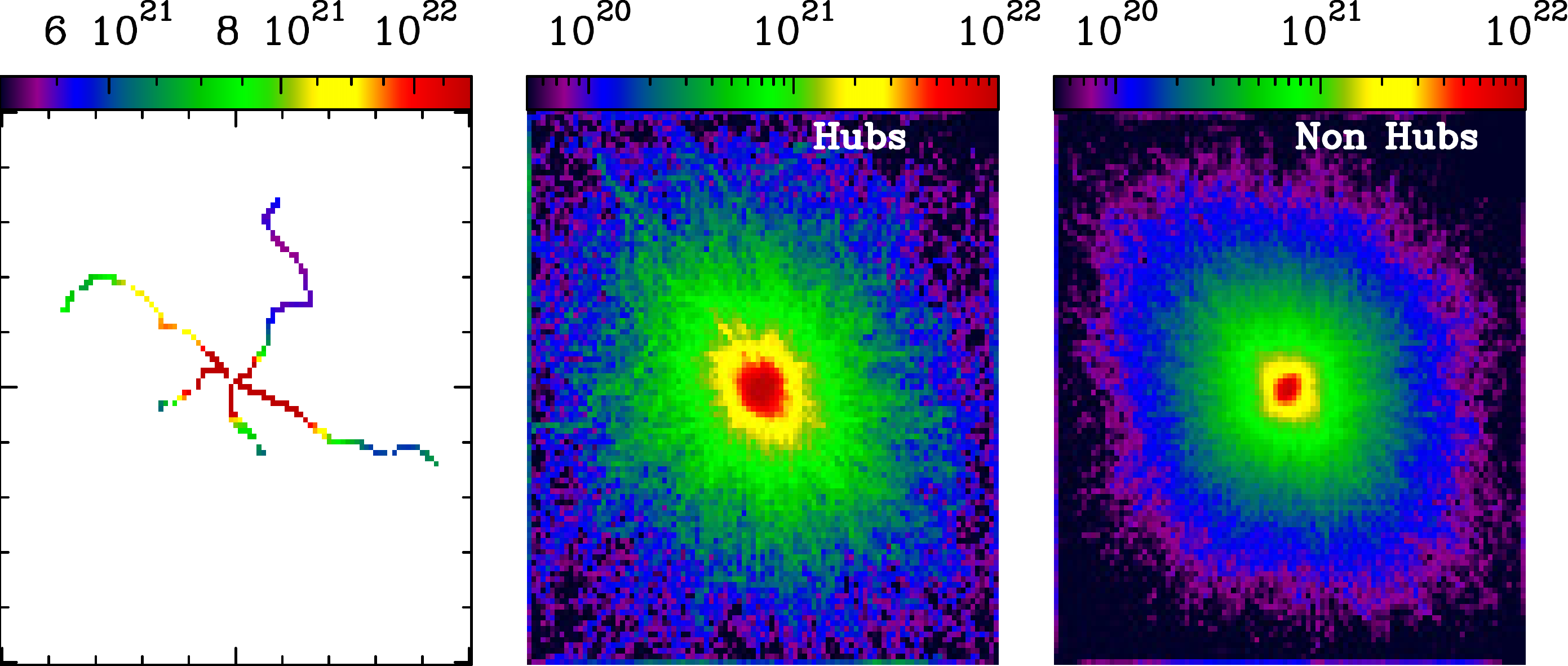}
   \caption{Computing the column density profile for HFSs: The column
     density map of each target is masked by the skeleton (left panel)
     and then averaged for all target hubs (middle panel) and
     non-hubs (right panel). The radial profile shown in Fig.\,11 is
     computed using these average images for different evolutionary
     groups. The example shown here is for the group of proto-stellar
     hubs and non-hubs, and the units of the colour bar are
     cm$^{-2}$. }
              \label{}%
    \end{figure*}

\section{Results}

In searching the 34,575 Hi-GAL clumps with the methods described
above, we found 3704 HFSs. Of these, 144 are saturated sources, all of
which are identified as HFSs.  This implies that $\sim$10.7\% of the
Hi-GAL clumps are located at the line-of-sight junction of filamentary
structures. Other clumps which are located on a single filament or at
the junction of two filaments are called non-hubs in the
following. There are 26135 non-hub clumps, of which 10380 are located
at the junction of two filament skeletons and 15755 form the tip of a
single filament. Many of these latter 10380 non-hub clumps may simply
represent a clump that is actually located somewhere along a single
filament. There are also 4736 clumps that are not associated with any
filaments (0 skeletons) and are not included in further
analyses. Using the evolutionary state classification in the Hi-GAL
catalogue, we find that 156 (4\%), 2010 (54\%), and 1537 (41\%) of the
3704 HFSs are respectively classified as starless, pre-stellar, and
proto-stellar in nature. All 144 clumps that are saturated in one or
more of the \herschel bands are proto-stellar in nature and they are
among the most active sites of star formation.

Inspection of the skeleton overlays on the images reveals that the
filamentary structures detected here can represent either (a) cold
dense filaments such as infrared-dark clouds that are conducive to
star formation, or (b) emission nebulae and similar structures (ex:
edge of a dusty shell or bubble illuminated by a massive star) with
significant dust column density located in an already active star
forming region. The distinction between such targets will require
multi-wavelength examination. For example, in Fig.\,4 we display
Spitzer IRAC 8\mum images of two hubs demonstrating that a hub may
represent any combination of 8\mum dark and bright filamentary
structures.

\begin{figure}
   \includegraphics[width=0.47\textwidth]{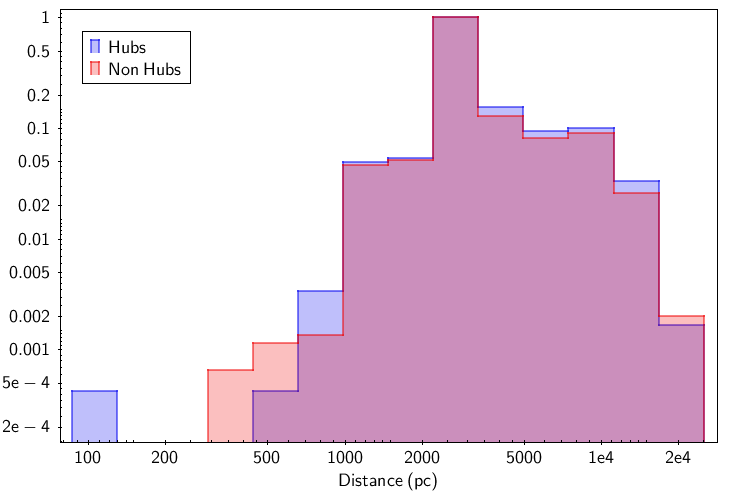}
   \caption{Histograms (normalised by maximum y-value) of distances of
     all clumps compared with those of HFSs, demonstrating that HFSs
     are uniformly distributed over the range covered by all clumps.
   }
              \label{}%
    \end{figure}
    
    \begin{figure}
   \includegraphics[width=0.47\textwidth]{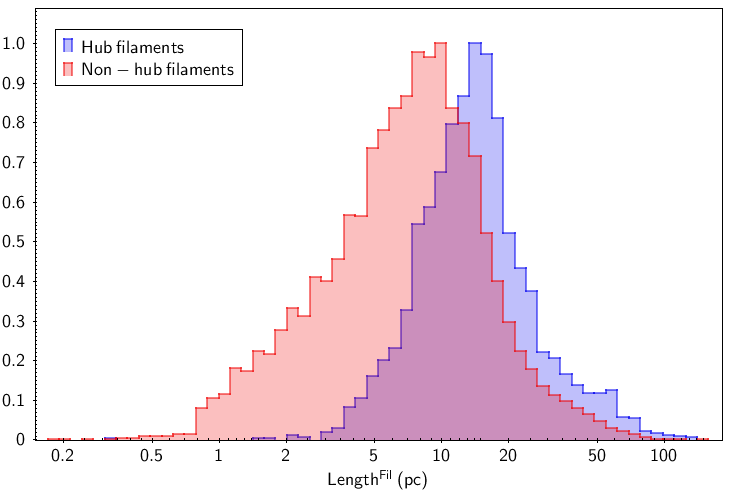}
   \caption{Histogram (normalised by maximum y-value) of filament
     lengths for HFS and non-HFS samples, showing that HFSs
     are composed of longer filaments. }
              \label{}%
    \end{figure}

    \begin{figure}
   \includegraphics[width=0.47\textwidth]{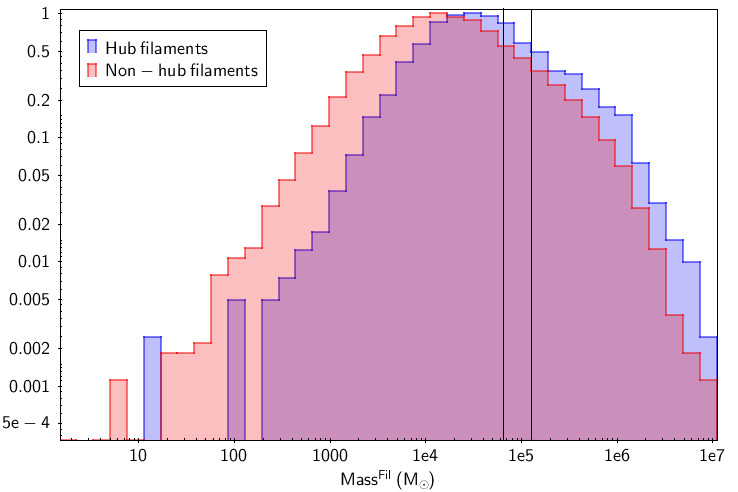}
   \caption{Histogram (normalised by maximum y-value) of filament
     masses in HFS and non-HFS samples, showing the effect of massive
     filaments in HFSs. {\bf The vertical lines represent the mean values.}  }
              \label{}%
    \end{figure}
   
    \begin{figure}
   \includegraphics[width=0.47\textwidth]{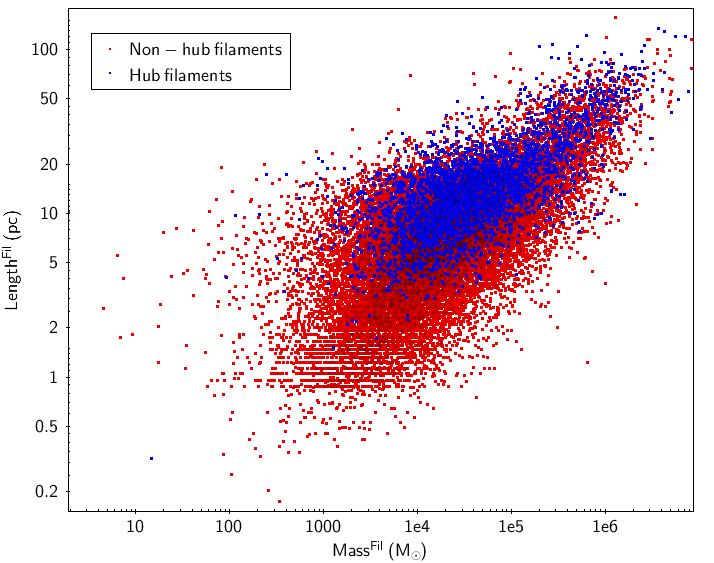}
   \caption{Mass--length relation of the filaments around clumps.}
              \label{}%
    \end{figure}

    \begin{figure}
   \includegraphics[width=0.47\textwidth]{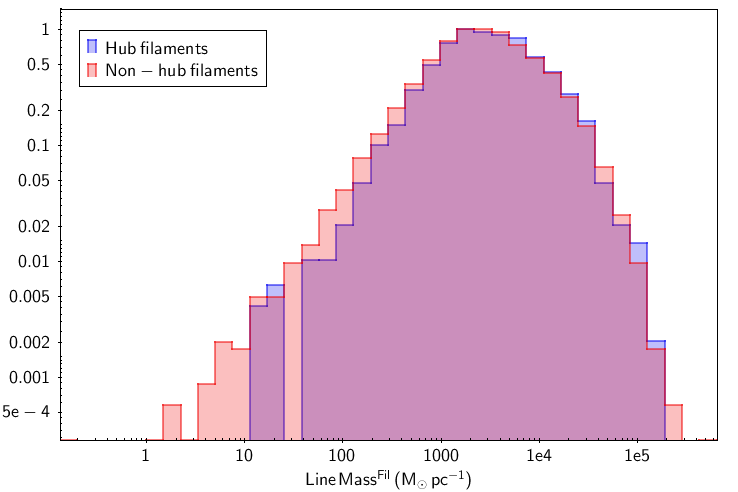}
   \caption{Histograms (normalised by maximum y-value) of filament line
     mass.  }
              \label{}%
    \end{figure}
We find that the identified HFSs are composed of up to seven filaments
joining at the hub. By examining the images of some of the brightest
sources, especially saturated sources, it is evident that there are
many more filamentary features visible on the images that have not
passed the series of selection criteria used in the detection method
described above. This is because a single uniform criterion that is
applied here cannot work at full efficiency for the variety of
targeted sources, especially with a complex mixture of brightness,
background, and structures. Persistence level and selection cuts have
to be appropriately fine-tuned to each target in order to extract all
features in a given image. Therefore, the upper limit on the number of
filamentary structures forming a hub is only moderately represented by
the n$_{skel}$ parameter in the HFS catalogue. This analysis however
yields a number of prominent skeletons (with S/N$>$5).

\subsection{Filament and hub properties} 

The targets in the studied sample have a range of distances between
100pc and 24kpc. The Hi-GAL clumps are generally recovered from major
areas of star formation, peaking roughly at 3\,kpc (Galactic fourth
quadrant, central bar), 5\,kpc (molecular ring, Sagittarius, and Norma
spiral arm tangents), and at 11\,kpc (Galactic warp) \citep[e.g. see
  Fig.\,4 in][]{Elia2017}. In Fig.\,6, the distance histograms
(normalised by the maximum value of each sample) of all the hubs and
non-hub clumps are shown, which indicate that the two are similarly
represented at all distances. Inclination angles of the HFSs will also
play a major role in their detection, especially at these large
distances; however, there is no measurement to distinguish those
effects. The 18.2\arcsec\, beam resolution of the 250\mum images
corresponds to the projected scales of 0.26pc and 0.7pc at distances
of 3kpc and 8kpc, respectively.  The filament widths are clearly
unresolved in the HFS sample; indeed the HFS detected filaments are
either high-line-mass filaments or elongated clouds, and are not
comparable to the filaments described by studies of nearby star
forming regions in the Gould belt \citep{Arzoumanian2019}. In Fig.\,7,
we plot the normalised histogram of filament lengths; this is taken as
length$^{out}_{fil}$ (in pc) instead of total filament length to
ensure that the fraction of the filament within the target clump FWHM
is excluded from consideration. We note that, `individual filaments'
in the following figures refers to the constituent filaments of
non-hub clumps. Figure\,7 shows that the hub systems are dominated by
longer filaments (mean length $\sim$18\,pc) compared to the non-hub
clump (mean length $\sim$8\,pc) average. Figure\,8 displays the
normalised histogram of filament masses, where one can see the
immediate effect of longer filaments influencing the mass. The mean
mass of the HFS filaments (1.4 $\times$10$^5$ \msun) is higher than
that of the non-hub clumps ($\sim$6.3$\times$10$^4$\msun). However,
this may not be significant given the broad distribution of the
masses. For a given sensitivity of the Hi-GAL maps, one of the most
profound effects of the limited angular resolution is that at larger
distances, only longer filaments can be detected. This is reflected in
Figs.\,7 and 8. The relation between filament length and mass can be
visualised in Fig.\,9; at lower masses and shorter lengths, this plot
displays a larger scatter when compared to similar plots from studies
of filament catalogues \citep{Li2016,Schisano2020}. This may be the
result of detecting filaments on the 250\mum images which has a higher
spatial resolution than the column density maps used in the other
studies. The striking feature of Figs.\,7, 8, and 9 is that even
though the candidate HFSs are uniformly distributed over the distance
range traced by the full target sample, HFSs appear to be composed of
longer and more massive filaments. It can also be viewed as lower
detection of weaker or less massive clumps at large
distances. Figure10 shows the normalised histogram of the filament
line mass, indicating a similar distribution for filaments
constituting both hub and non-hub systems.

\subsection{Density enhancement and massive star formation in hubs}

Circularly averaged radial profiles of column density centred on the
clumps located in hubs and non-hubs are shown in Fig.\,11. These
profiles are also separated for pre-stellar and proto-stellar
objects. Targets with saturated pixels are all proto-stellar but they
are shown separately to distinguish modelled fluxes using Gaussian
fitting from the rest. Moving along the filaments to the clump centre,
the column densities of the clumps representing hubs display
enhancements when compared to those located in non-hub systems;
i.e. clumps located on individual filaments. The ratio of the peak
column density at the centre of the clump between the hub and non-hub
systems is taken as the enhancement factor which is 1.9 for
pre-stellar clumps, 2.1 for proto-stellar sources, and $\sim$10 in
saturated sources. It should be noted that most nearby regions of
intense star formation, producing the high-luminosity sources, are all
saturated in one or more bands.

\begin{figure}
 \includegraphics[width=0.47\textwidth]{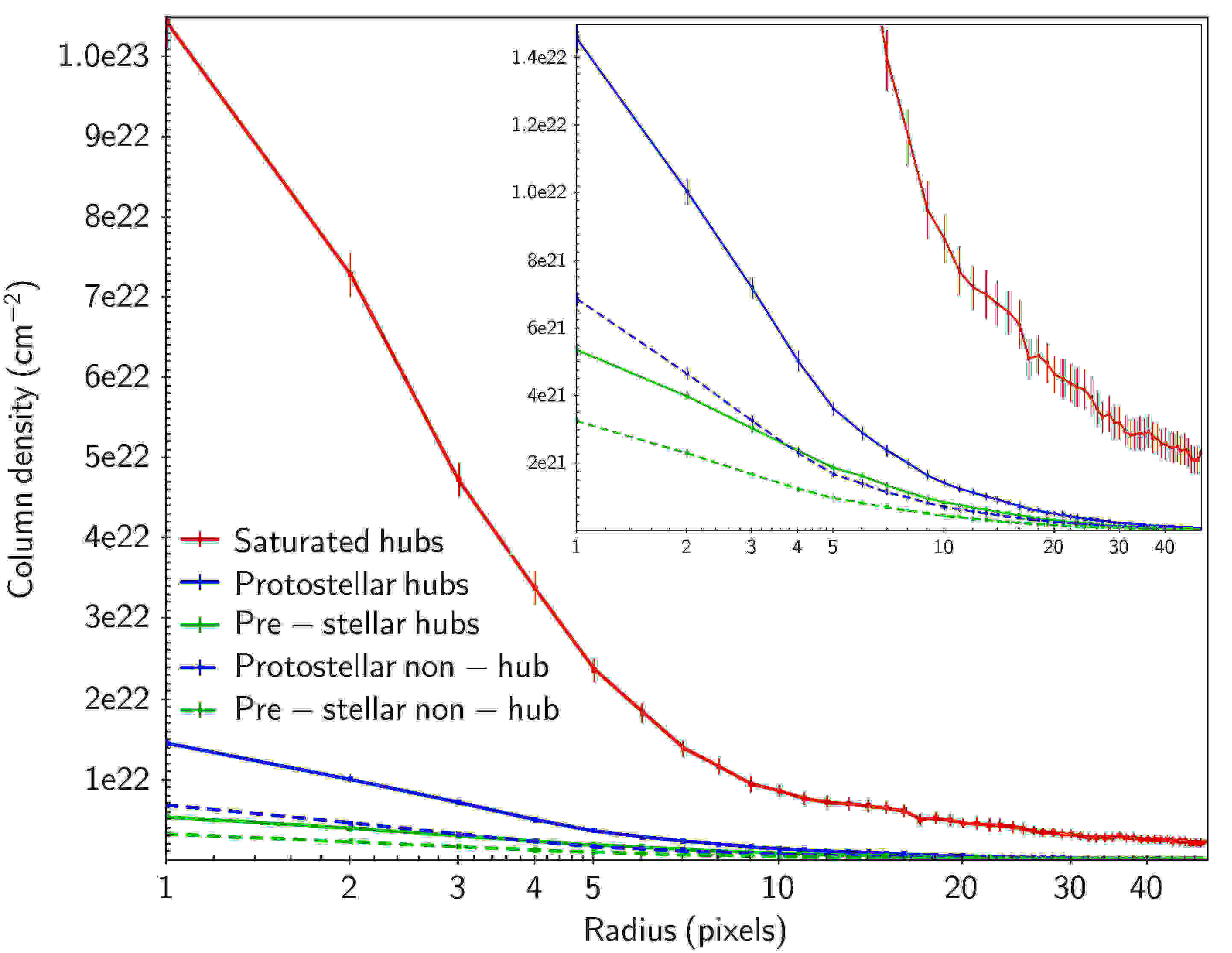}
\caption[]{ Circularly averaged radial column density profiles centred
  on the studied clumps. The column density is plotted as a
    function of pixel (each pixel is an azimuthal average) distance
    from the centre. The error bars display the standard deviation on the
    azimuthal average at each pixel. Saturated proto-stellar clumps
  are plotted separately because the fluxes are modelled and recovered
  from Gaussian fitting.}

\label{fig:radplot}
\end{figure}

\begin{figure}
\includegraphics[width=0.47\textwidth]{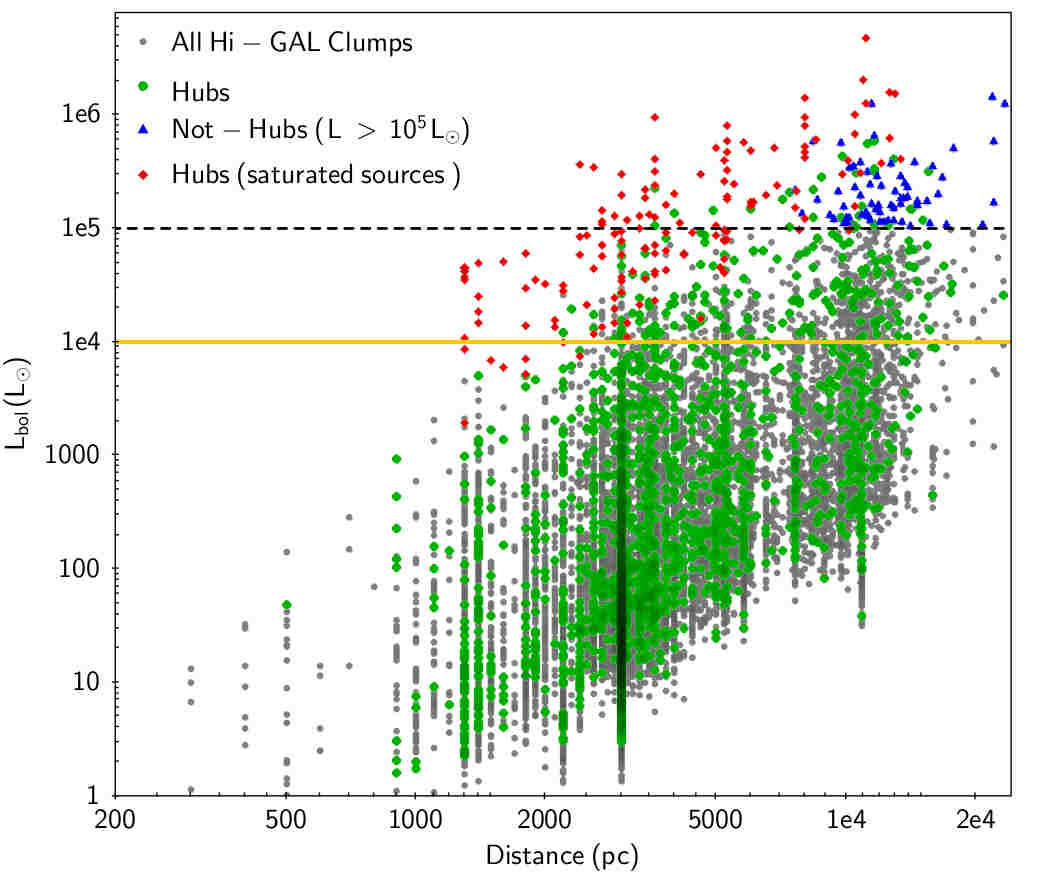}
\caption[]{All massive stars form in hubs: All sources with a
  luminosity L$\ge$10$^5$\lsun in the inner Milky Way and located at a
  distance of $<$5\,kpc are found to be HFSs. All hubs are marked by
  green circles. Clumps (with L$\ge$10$^5$\lsun) that are not found to
  be hubs (blue triangles) are located farther than 5\,kpc because the
  18\arcsec\, beam resolution of the data is insufficient to resolve
  these structures. The yellow line at L = 10$^4$\lsun indicates the
  Eddington ratio at which radiation and gravitational pressures
  become roughly equal; at $<$2\,kpc, all objects above this
  luminosity are hubs, demonstrating that massive stars preferentially
  form in hubs.}
\label{fig:corollary}
\end{figure}


\begin{figure*}
\includegraphics[width=0.97\textwidth]{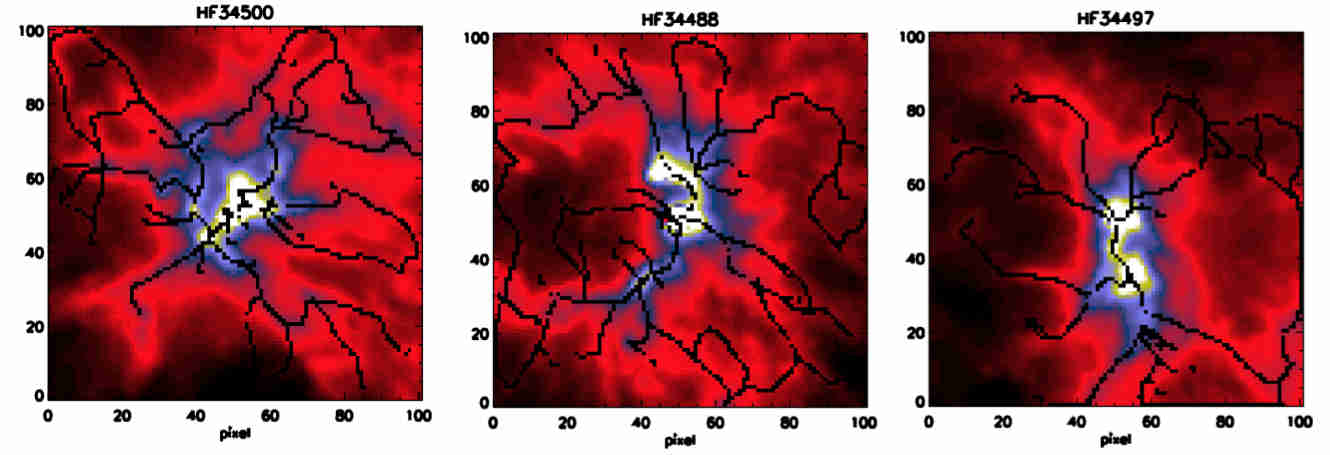}
\caption[]{Examples of saturated sources with a large network of
  filaments. Skeletons are overlaid on 250\mum images.}
\label{fig:corollary}
\end{figure*}

The distribution of hub and non-hub systems on a luminosity vs.
distance plot is shown in Fig.\,12. Saturated sources are plotted
separately and are all proto-stellar in nature. The two horizontal
lines mark the luminosity cuts at 10$^4$\lsun and 10$^5$\lsun. All
clumps with L$>$10$^{5}$\lsun located at a distance of $\le$5\,kpc are
hubs, as are the clumps with L$>$10$^{4}$\lsun at $\le$2\,kpc. At
farther distances, identifying the HFSs by resolving individual
filaments is limited by the 18\arcsec\, resolution of the 250\mum\,
images. Stars of masses greater than 8\msun are considered massive;
however, only in stars with a luminosity L$>$10$^4$\lsun \ do the
radiation and gravitational pressures become roughly equal. Given that
essentially all luminous targets at nearby distances are hubs, this
means that all massive stars form in hubs, especially those where
radiation pressure is considered significant, defined by the Eddington
ratio of one.

The results shown in Figs.\,11\, and 12 lead us to the following
corollary;{ the most natural conditions to form the massive stars
  arise only when multiple filaments join to form a hub, because this
  is when the characteristic densities (and mass) of the individual
  filaments can be instantaneously summed, creating the highest
  density and most massive pockets of gas and dust.} The coalescence
scenario by \citet{BBZ1998} proposed that low-mass stars merge and/or
coalesce to form high-mass stars in a high-density environment. In its
essence, the corollary above is the coalescence scenario, except that
the merging is not of the stars, but of the gas and dust in the
fertile \citep{Hacar2018} filaments.

For majority of the clumps studied (unsaturated in the \herschel
bands), the number of skeletons joining at the hub is in the range of
3-7. On the contrary, the saturated sources display large networks of
filaments around them, typically 6-12 main filaments. A comparison of
the skeleton numbers between saturated and unsaturated clumps is not
appropriate because we do not clip the filaments in saturated sources
using column density cuts (see Sect. 3.2). Also, filament detection
must be improved using 160$\mum$ and 70$\mum$ data along with that of
250\mum\ in order to enhance spatial resolution. However, it should be
mentioned that in the saturated sources, which represent the most
luminous and nearby regions and are bright sources, the column density
is high throughout the field. In Fig.\,13, we show samples of
saturated sources along with the skeletons. These mostly represent
major nodes of clustered star formation in giant molecular clouds,
such as the ONC. The result shown in Fig.\,11 is evidence that HFSs
with the largest density increase at the hub (saturated sources) are
the result of larger networks of criss-crossing filaments.

\section{Filaments to clusters: A paradigm for star formation}

These new findings from Sect. 4.2 lead us to build a scenario of star
formation in the HFS paradigm as presented in Fig.\,14. This scenario
is represented by four stages that can be roughly compared with
evolutionary snapshots of star formation in molecular clouds. Stages
I, II, III, and IV respectively represent low-mass-star formation in
filaments without hubs, pre-stellar HFSs surrounded by young clusters
of low-mass stars, HFSs with high-mass protostars surrounded by young
cluster of low-mass stars, and a full-blown HII region with embedded
clusters such as the Orion Nebular Cluster (ONC).

Low- to intermediate-mass star formation alone can take place in
individual filaments, whereas high-mass stars form preferentially in
hubs. In the following, we elaborate on this scenario using the
simplest case of two filaments coming together at a junction and thus
forming a hub.

\begin{figure*}
\includegraphics[width=0.95\textwidth]{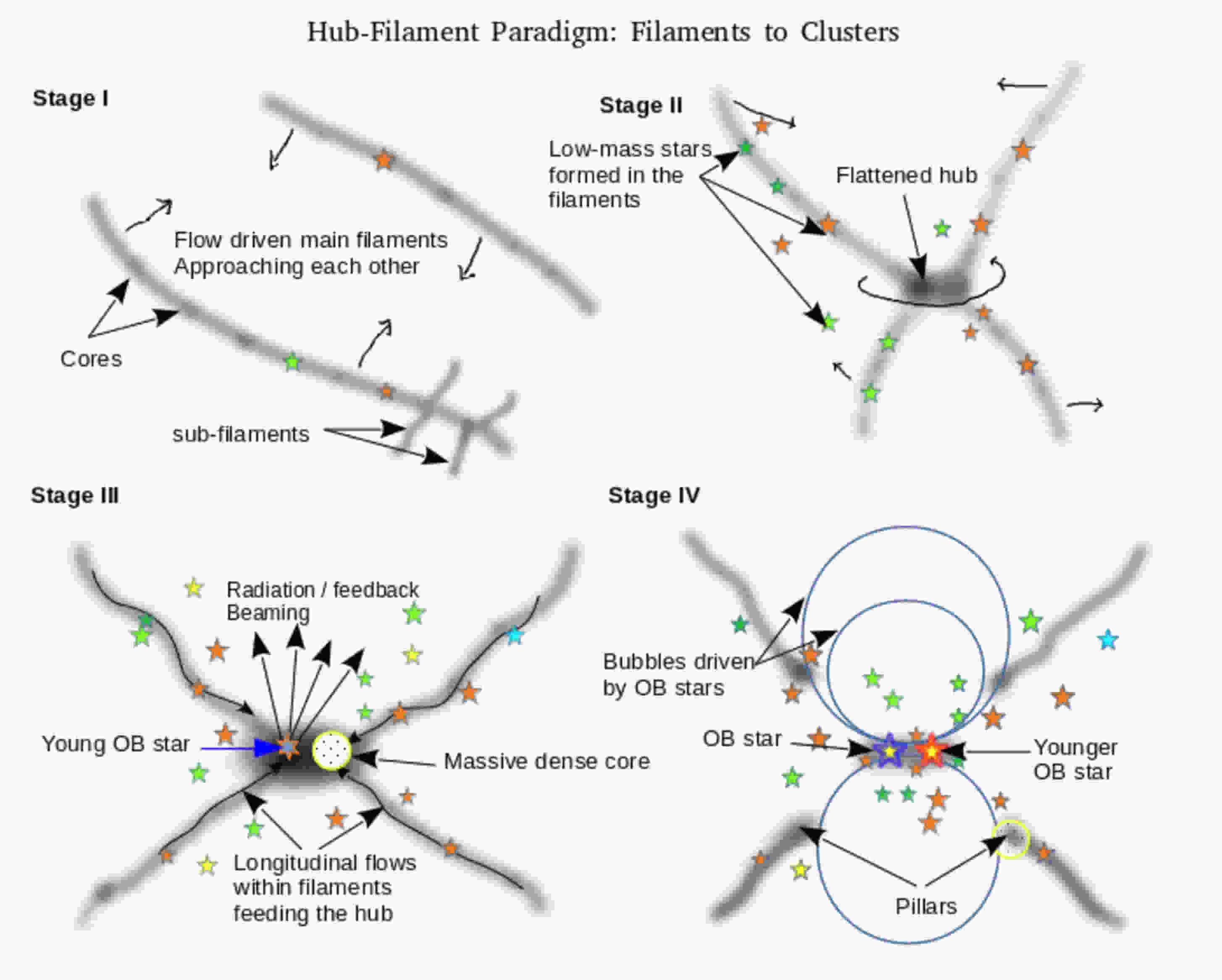}
\caption[]{{\bf Filaments-to-clusters paradigm for star formation.}
  Flow-driven filaments overlap to form a junction that is called a
  hub. The hub gains a twist as the overlap point is different from
  the centre of mass, and the density is enhanced due to the addition
  of filament densities. Low-mass stars can be ongoing before and
  during hub formation. The hub gravitational potential triggers and
  drives longitudinal flows bringing additional matter and further
  enhancing the density. Hub fragmentation results in a small cluster
  of stars; however, a pancake or sheet morphology often leads to
  near-equal mass fragmentation, especially under the influence of
  magnetic fields and radiative heating. Radiation pressure and
  ionisation feedback escapes through the inter-filamentary cavities
  by punching holes in the flattened hub. Finally, the expanding
  radiation bubbles can create bipolar shaped HII regions, burning out
  the composing filaments to produce tips that may be similar to the
  structures called pillars. The net result is a mass-segregated
  embedded cluster, with a mass function that is the sum of stars
  continuously created in the filaments and the massive stars formed
  in the hub.}

\label{fig:paradigm}
\end{figure*}

\vskip 2mm
\subsection{Stage I: Initial conditions}
Flow-driven filaments, either due to intra-molecular cloud velocity
dispersions ($\sim$1~km~s$^{-1}$) or externally driven by expanding
shells ($\sim$10~km~s$^{-1}$) join to form a hub. Observations of
colliding filaments in the Mon OB1 star-forming region show a pair of
filaments approaching each other with relative velocities of
2-4~km~s$^{-1}$ \citep{Montillaud2019}. A variety of external triggers
can also drive such motions, of which stellar wind bubbles and HII
region shells are observationally prominent \citep[see][for a
  discussion]{Myers2009}; but late phases of supernova remnants are
also important. Such filaments should be similar to the filaments in
Taurus \citep{Palmeirim2013} and are often fertile \citep{Hacar2018},
with ongoing low-mass star formation. This is because large
populations of young low-mass stars have been found around high-mass
protostars \citep{Kumar2006} and around HFSs
\citep{Dewangan2017,Baug2018}.

The filaments that set up the initial conditions for the formation of
the hub need not necessarily form by gravity and turbulence. Instead,
they can form via mechanisms such as cloud--cloud collision where the
most basic effect of compression will lead to filament
formation. Numerical simulations by \citet{Inoue2013} describe the
formation of such filaments and also show that they are magnetised
\citep[see also][]{Inoue2018}, which is an important aspect in the
formation of massive stars as is seen at Stage III. Numerous
observational studies \citep[e.g.][]{Fukui2014,Torii2015,Sano2018} are
in support of this mechanism. Indeed this may be the prominent
mechanism through which hubs with large networks of filaments can form
in the most massive clouds, leading to cluster formation. In summary,
dense fertile filaments moving towards each other set up the initial
conditions of the HFS paradigm.


\subsection{Stage II: Hub formation -- spin and geometry}
The filaments can overlap with any relative orientation (0-90\deg), in
such a way that a pre-existing dense core, intra-filamentary material,
or a combination of both can form the junction. At small inclination
angles, multiple filaments are compressed essentially along the
longitudinal axis; this may be the mechanism responsible for the
formation of high line-mass \citep{Kainulainen2017}, and a large
network of connected filaments \citep{Hacar2018} such as the Orion
integral shaped filament. This would not constitute a hub, even though
high-line-mass filaments may form on an average higher-mass cores than
their lower mass counterparts \citep{Shimajiri2019,Konyves2019}. A hub
is a relatively low-aspect-ratio object \citep{Myers2009} and together
with its density amplification property (Sect. 4.2) it will always
provide enhanced star formation conditions with respect to what is
possible within a filament of certain line mass.  In other words, a
hub can form a massive star that is more massive than the most massive
star that can form in the hub-composing filaments.  The larger number
of filaments in saturated sources together with the density
amplification result prompts us to predict that the mass of the most
massive star formed will be correlated with the network factor
f$_{net}$ = $\sum_{fil=1}^{n} N_{fil}^{M_{line}} \times
M_{line}^{fil}$ where N$_{fil}^{M_{line}}$ represents the number of
filaments with a certain line mass M$_{line}^{fil}$.

The probability to overlap at a location that is not exactly the
centre of mass is high. Therefore, we propose that the approaching
filaments can impart a small twist to the hub, giving rise to an
initial angular momentum. If the colliding filaments have large
line-mass inhomogeneity prior to the collision, the twist and
therefore the rotation of the hub is also larger. We suggest that the
resulting spin can eventually flatten the hub, a conjecture that can
be observationally examined. This may also explain the new ALMA
observations of MonR2 \citep{TrevinoMorales2019} where a spectacular
network of filaments can be seen spiralling into the
hub. \citet{Myers2009} compared the observed HFSs in nearby regions of
star formation to several analytical models and argued that the outer
layers are best represented by a modulated Schmid-Burgk
equilibrium. He focussed on explaining the formation of hubs and
parallel spaced filaments in nearby regions and arrived at a pancake
or sheet-like geometry for hubs. Numerical simulations of cloud
collision and compression \citep{Semadeni2007} produce a central shell
with outwardly radiating filaments. By any or all of these arguments,
a hub is very likely to possess a flattened geometry.

Indeed, a hub is likely to assume an ellipsoidal geometry, and more
specifically an oblate spheroid. The mean aspect ratio of hubs from
our candidate sample, measured using the 250\mum images, is
1.2$\pm$0.4, however $\sim$20\% of them have an ellipticity of
$\sim$1.5--3.0. The 18\arcsec\, angular resolution of the data is
insufficient to resolve the ellipticity for all targets, especially
when considering projection and distance effects, but it is evident in
the high-resolution data of individual targets \citep[][see also
  Fig.\,4]{Williams2018,Chen2019}.

A hub with an oblate spheroidal geometry is pre-stellar at this stage,
and observationally represented by a massive pre-stellar clump
surrounded by a population of low-mass young stellar objects
(YSOs). An excellent example representing this stage is the M17SWex,
where the infrared dark cloud (IRDC) display approximately 500 YSOs
(all low mass, conspicuously lacking massive objects) in the dark
cloud \citep{Povich2010} containing at least two hubs
\citep{Chen2019}.

In the simplest case of two, typically fertile filaments, the overlap
can happen in such a way that a pre-existing dense core,
intra-filamentary material, or a combination of both can form the
hub. Therefore, a hub will inherit the combined density
inhomogeneities of the composing filaments, as depicted in Fig.\,14 by
the relatively dense left edge of the hub compared to the right at
this stage.

\subsection{Stage III: Massive star formation in the density amplified hub} 
A hub formed by the junction of filaments is an object experiencing
shock. \citet{Whitworth1994} argue that a suddenly compressed layer
will switch to become confined by self-gravity, moving from a flat
density profile to a centrally condensed density over a time
t$_{switch}$. This time is of the order of one free-fall time of the
uncompressed medium. If we consider the dense gas in the individual
filaments as the uncompressed medium, a hub should become
gravitationally unstable in $\sim$1-2\,Myr after its formation. This
time is also similar to the estimated mass doubling time of 1-2\,Myr
in filaments \citep{Palmeirim2013}. However, a magnetically threaded
hub can take longer to begin to collapse.

Having inherited density inhomogeneities at Stage II, the densest
portion of the hub begins to collapse first, forming the first massive
star. The remaining portion will collapse subsequently, leading to the
second most massive object, which should be relatively young compared
to the previous massive star. If further fragmentation were to happen
at each of these centres, one may expect two groups of objects with
relatively different evolutionary states.  Observations of massive
stars in HFSs often show two formation sites within a hub;
interestingly, these can have similar masses (luminosity) and slightly
different ages, with one older (IR-bright) and one younger (IR dark
and sub-mm bright). Some examples can be found in Fig.\,4a, SDC13
\citep{Peretto2013,Peretto2014,Williams2018}, and in Fig.\,1 of
\citet{Kumar2016}. The trapezium cluster and the BN/KL object
represent such a pair in the ONC identified by the two dust
condensations as the hub \citep{Myers2009}. N$_2$H$^+$ observations of
the same region show the highest density of fibres centred on the
BN/KL object where high-mass star formation is ongoing. In contrast,
the molecular gas is evacuated around the trapezium cluster. Nearby
young clusters also display such a pattern, where one focal point of
an elliptical shaped cluster is more evolved than the other
\citep{Schmeja2008}. Therefore, we suggest that hubs with their oblate
spheroidal geometry tend to form two near-equal massive objects or
groups of objects, with a relative difference in evolutionary
state. Interestingly, non-axisymmetric numerical simulations of
high-mass star formation presented by \citet{Krumholz2009} produce a
near equal-mass pair of high-mass stars with a time difference.
 
The column density amplification in the hub produces a gravitational
potential difference between the hub and the filament, which can
trigger and drive a longitudinal flow within the filament (analogous
to electric current in a wire) directed toward the hub. Such flows
were observed in the SDC13 massive star forming HFS
\citep{Peretto2013,Peretto2014,Williams2018} and recently in the
M17SWex region \citep{Chen2019}. The mass flow rates of
10$^{-4}$-10$^{-3}$\msun yr$^{-1}$ reported by \citet{Chen2019} are by
themselves sufficient to form massive stars. We suggest that in this
way, the filaments act as the secondary reservoir feeding the hub
(primary reservoir) to sustain its density conditions, so that the
central region of the hub does not run out of gas; this is necessary
because if it does, the massive protostar will stop burning hydrogen
and begin to accumulate helium ash, moving away from the main
sequence.

While the hubs may be responsible for the initiation of a longitudinal
flow, the flow in turn can trigger gravitational collapse in a stable
hub. The enhanced density conditions in the hub may be comparable to
that of monolithic dense cores required by the core accretion models
\citep{McKeeTan2003}, but fragmentation is the key issue that sets
constraints on the formation of the highest mass stars
\citep{Peters2010,Krumholz2015}. Magnetic fields can offer the
stability against collapse and fragmentation. B-field observations
\citep{Wang2019,Beltran2019} of HFSs attribute nearly equal importance
to gravity and magnetic fields, and less importance to turbulence. {We
  propose that the flow of matter and the density increase in the hub
  can be expected to compress the initial/local magnetic field,
  thereby increasing its strength and stabilising the hub against
  multiple fragmentation into low-mass objects, and favour fewer
  fragments of comparable (high) mass \citep{MyersMcKee2013}. } Even
if the hub fragments into lower mass objects, these objects cluster
can grow to become a cluster of higher mass stars while deriving high
accretion rates via longitudinal flows \citep{Chen2019}. It has been
suggested that Bondi-Hoyle accretion can be increased if the virial
parameter is small, matter flows onto a cluster of stars
\citep{KetoWood2006}, and/or if the infall originates from a
significantly less massive clump. In such a scenario, the group of
fragments in the hub and the longitudinal flows along the filaments
respectively represent the cluster of stars and infall from a less
significant reservoir. Indeed, this may be the scenario witnessed by
old VLA observations \citep{Keto2002} used to estimate accretion rates
of 10$^{-3}$\msun yr$^{-1}$ and also new ALMA observations of Mon R2
\citep{TrevinoMorales2019}, where a spectacular network of filaments
are seen spiralling into the cluster centre. Virial properties of
massive pre-stellar clumps are shown to be dominated by gravity rather
than turbulence \citep{Traficante2020} and are shown to decrease with
size-scale at least in one case \citep{Chen2019}. Therefore, one might
expect hubs to be far less prone to turbulent fragmentation, and thus
more prone to forming the highest mass ($>$100\msun) stars.

\subsection{Stage IV: Embedded cluster and HII regions}
 \citet{Arzoumanian2019} show that $\sim$15\% of the total cloud mass
 is dense gas, 80\% of which is in the form of filaments. These
 authors estimate that the area filling factor of filaments \citep[see
   their Table.\,1; see also][]{Roy2019} is only 7\%, which in turn
 indicates a very low volume filling factor. Therefore, HFSs offer
 natural structural vents to efficiently beam-out the radiation
 pressure to the inter-filamentary void, by punching holes in the
 flattened and centrally located hub. The same holes can eventually
 serve to dispel a significant portion of ionised gas pressure and
 stellar winds. Numerical simulations \citep{Dale2011} of
 massive-star-cluster formation clearly display this effect. It is
 argued that this mechanism is responsible for not eroding the
 molecular filaments, instead filling the inter-filamentary voids and
 bubbles with ionised gas. An assessment of the feedback factors in
 HII regions \cite{Lopez2014} shows that both direct radiation and hot
 gas pressures leak significantly. Only the dust-processed radiation
 and warm ionised gas pressures are found to impact the HII
 shells. The flattened geometry of the hub allow holes to form along
 the minor axis of the oblate spheroid, or the sheet. We propose that
 massive star formation in the hubs is responsible for the observed
 bipolar HII regions. Also, the bubbles of radiation and ionised gas
 eating through the parent filaments may be forming what is known as
 `pillars of creation' found in, for example, the Eagle Nebula.

Many HII regions display bipolar morphology
\citep{Deharveng2015,Samal2018}, and a large number of bubbles are
catalogued in the Milky Way \citep{Palmeirim2017}, some of which are
thought to be bipolar HII regions viewed pole-on. Bipolar HII regions
are found to be driven by massive stars forming in dense and flat
structures that contain filaments \citep{Deharveng2015}. Our
proposition above in the HFS scenario is well represented by these
observations in terms of morphology, stellar population, and
mass-segregation. \citet{Whitworth2018} suggest cloud--cloud collision
as a mechanism to produce bipolar HII regions and massive star
formation; however, this scenario is only valid when assuming a
spherical geometry for the clouds. Given the ubiquity of the
filamentary nature of the clouds, our proposition here in the HFS
scenario better reflects the observations. Magnetic field observations
of bipolar HII regions \citep{Eswaraiah2017} display an hourglass
morphology closely following the bipolar bubble. The field strength
itself suggests a magnetic pressure dominating the turbulent and
thermal pressures. This is consistent with the arguments made for
Stage\,III, prompting further observational investigation of the role
of magnetic fields at earlier stages.

\subsection{Salient features of star formation in HFSs}

{\bf\em Mass segregation:} At the completion of star formation and gas
dispersal after a few million years, the HFS would have led to the
formation of a mass-segregated young cluster. In general, the central
location of the hub in the HFS naturally leads to mass
segregation. An interesting discussion on the completely
  mass-segregated nature of Serpens south and ONC clusters can be
  found in \citet{Pavlik2019}. 
\vskip 3mm

\noindent{\bf \em Formation times and sequence of star formation:
}Based on the arguments made for Stage III, we propose that star
formation takes place quickly ($\sim$10$^5$\,yr) in the hub, forming a
group of OB stars with a top-heavy mass function, while lower mass
stars would begin to form in the individual filaments even before
assembling the hub and proceed slowly ($\sim$10$^6$\,yr). The
resulting sequence of the star formation by which low-mass stars form
prior to high-mass stars ensures that the low-mass star formation is
not adversely affected by the feedback from massive stars while
forming in a common environment, yet reconciles with the different
formation timescales. This sequence is quite evident in observations
of almost all nearby massive star forming regions, such as Orion
Nebula Cluster, MonR2, Lagoon Nebula, and so on, where the massive
star formation is ongoing at the centre, surrounded by dense clusters
of young low-mass stars. Studies of clustering around massive
proto-stellar candidates
\citep{Kumar2004,Kumar2006,Kumar2007,Ojha2010} demonstrate that the
sequence begins to take effect from the very early stages.

\vskip 3mm
\noindent{\bf \em Top-heavy initial mass function (IMF) in hubs and
  bound clusters:} In the HFS paradigm, hubs are where the massive
stars form, which can therefore lead to a stellar association that
will display a top-heavy IMF. If one were to measure the IMF in the
ONC, by considering stars enclosed within contours of different
stellar density, such as that of Fig. 3 in \citet{Hillenbrand1998}
encompassing the trapezium cluster, the mass function measured in the
inner most contour will naturally be top-heavy. This can be visualised
for example considering only the high-mass end of the Trapezium
cluster mass function \citep{Muench2002, LadaLada2003}. The Trapezium
cluster measures roughly 0.3-0.5\,pc which would be similar to the
size of a hub. When considering the stellar population averaged over a
longer timescale and larger spatial scale to encompass the giant
molecular clouds, a Salpeter slope can be effectively extended to the
higher mass end.

Hub-filament systems leading to the formation of even higher mass
stars than that found in the Trapezium cluster, and therefore bound
clusters, arise only at the junctions of multiple, high-line-mass
filaments. Combining the effects of angular momentum described in
StageII, such hubs can likely result in swirling spiral arm patterns
as evidenced in the MonR2 region \citep{TrevinoMorales2019}. Examples
of very high mass HFSs are star forming regions such as W51, W43, and
so on.  Recently, there was an interesting claim of a top-heavy CMF in
W43 \citep{Motte2018Nat}, where the region traced may represent the
main hub of that target. Other observations of HFSs with ALMA
\citep{Henshaw2017} indicate that hubs can form a spectrum of low- and
high-mass objects. However, from our arguments for Stage III and IV,
such a spectrum of objects may simply represent the seeds that can
grow further by mass accretion through longitudinal flows. The result
will be an association of OB stars, especially when the lower mass
stars grow to become intermediate-mass stars.

\noindent{\bf \em Age spreads in bound clusters:} When a hub is
composed of a large network of filaments, the individual filaments may
be drawn from a sample of fertile filaments that represent a wide
range of evolutionary timescales and star formation histories. Based
on an estimate of core life times that reside within filaments,
\citet{Andre2014} suggest a minimum of 10$^6$yr as the lifetime of
filaments. In the absence of massive star formation, which will clear
off the molecular gas in about $\sim$3-5 Myr, molecular cloud
lifetimes can be as long as 10 Myr or more \citep{Inutsuka2015},
during which time filament formation and destruction may happen.
\citet{LadaLA2010} suggest that, at least in the star forming regions
such as Ophiuchus, Pipe, Taurus, Perseus, and Lupus, the star
formation timescale is $\sim$2Myr and has not led to modification of
the total mass of the high-extinction material within the clouds. This
implies that low-mass-star formation can take place inside dense
fertile filaments without destroying them. Therefore, if star
formation began in the individual filaments at different times before
joining to form the hub, it can result in age spreads of a few million
years for low-mass stars. Once the hub forms, it can accrete several
smaller infertile filaments. The overall age spread of stars in a
cluster can therefore result from the oldest fertile filament to the
last star forming core in the hub. An interesting alternative
  explanation for age spreads can be found in \citet{Kroupa2018}.

\begin{figure}
\includegraphics[width=0.49\textwidth]{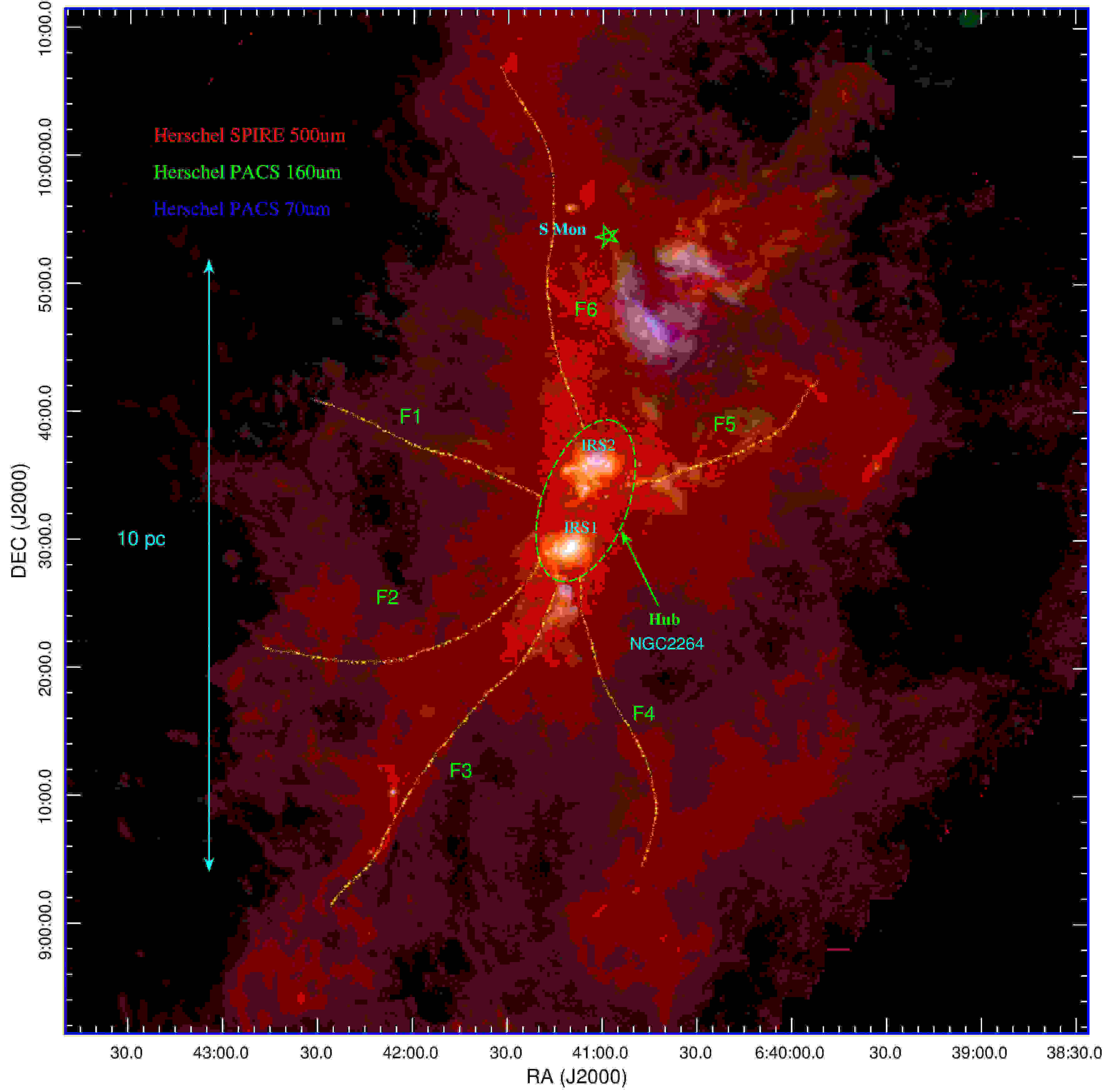}
\caption[]{NGC2264 as an example of StageIII: This is a HFS spanning
  $\sim$10pc, where IRS\,1 (above cone nebula) and IRS\,2 (centre of
  spokes cluster) represent the two eyes of the hub shown by the green
  ellipse.}
\label{fig:paradigm1}
\end{figure}

\begin{figure*}
  \centering
\includegraphics[width=0.8\textwidth]{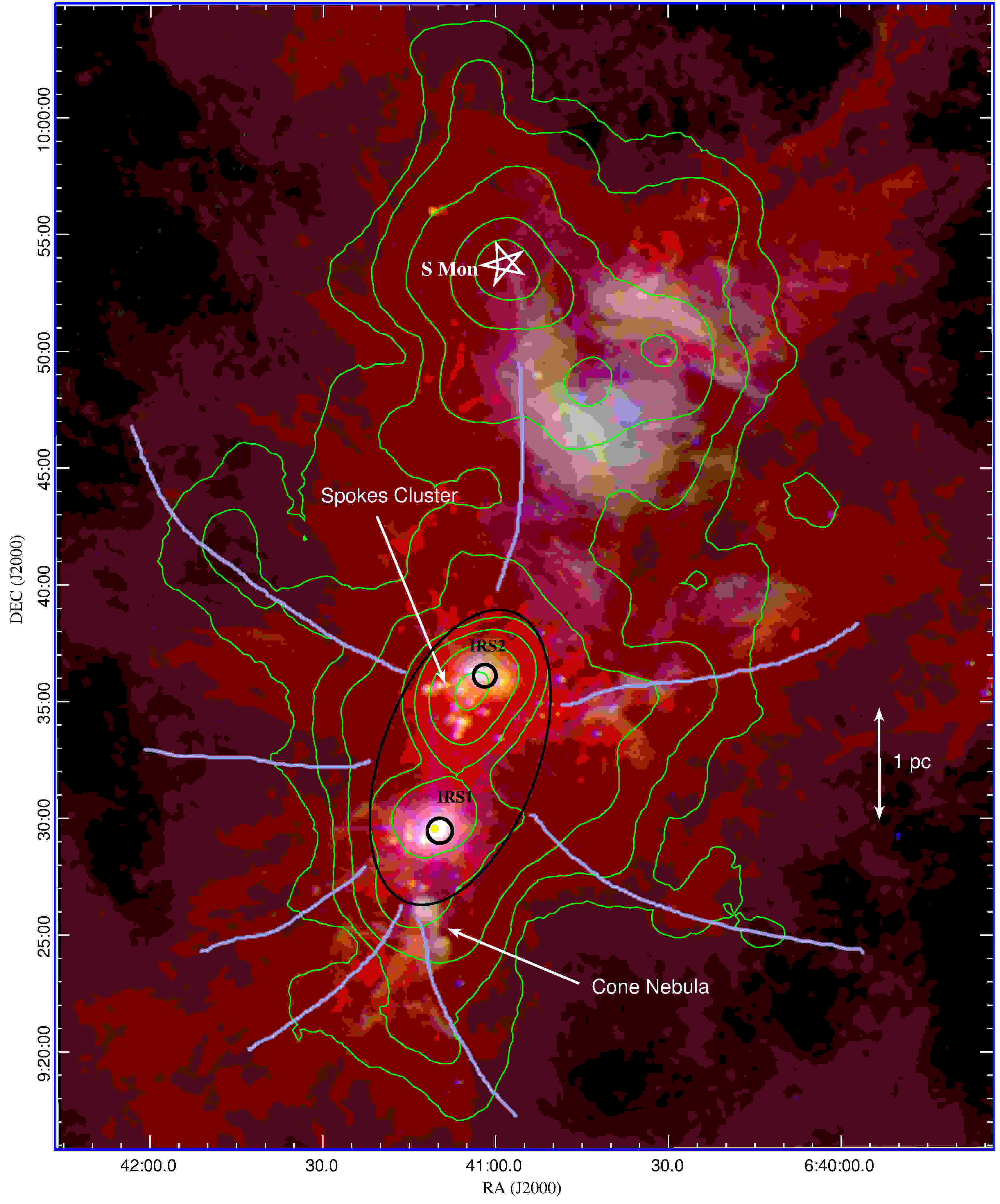}
\caption[]{Filaments of young stars in NGC2264: Evidence that the
  filamentary structure of the gas clouds is inherited by the density
  distribution of young stars as shown by the green contours (tenth
  nearest neighbour density of cluster members catalogued by
  \citet{Sung2009}). The violet line sketches mark these filaments of
  young stars. The hub has two centres of near-equal-intensity star
  formation with skewed evolutionary states. IRS1 is a $\sim$10\msun
  B2-type star well-known as Allen's source, and sits above the cone
  nebula (a pillar irradiated by this star). IRS2 is the younger
  150\lsun Class I type star surrounded by linearly aligned (in
  projection) young stars (Spokes cluster) detected at shorter than
  70\mum. The YSOs in the Spokes-cluster display at least three
  distinct groups of radial velocities that may correspond to the
  filaments in which they formed. The colour-composite image is made by
  combining images from SPIRE 500\mum (red), PACS 70\mum (green), and
  MIPS 24\mum (blue) data.}
\label{fig:paradigm2}
\end{figure*}

\noindent{\bf\em OB associations:} The star formation scenario in HFSs
can also explain the formation of isolated OB
associations. \citet{Ward2019} used Gaia-DR2 data to argue that the
kinematic properties of at least some OB associations strongly suggest
in-situ formation, and not the remnant of a dissolved cluster. Here we
propose that such associations can be the product of star formation in
the hubs, where the hub-composing filaments were not fertile enough to
initiate significant low-mass star formation in the individual
filaments, and that the mass in the individual filaments was fed to
the hub via longitudinal flows to form a group of higher-mass stars
resulting in the OB associations. Therefore, we suggest that hubs that
become gravitationally unstable before the individual filaments can
form a significant population of low-mass stars that can lead to the
formation of OB associations without associated low-mass clusters.

\section{Comparison of the HFS paradigm with NGC2264 and W40}

The ideas and arguments presented in the previous section are best
described by two well-studied star forming regions located within
1\,kpc of the Sun. NGC2264 in the Monocerous region and W40 in the
Aquila rift represent StageIII and StageIV of the HFS paradigm,
respectively. A large number of previous observations and publications
exist for these targets, especially the NGC2264 region. In the
following, we revisit some of the literature and demonstrate how the
published data reflect, and are sometimes better explained when viewed
in, the HFS paradigm.  We caution that the arguments made below do not
necessarily reflect the interpretations of the original literature.

\subsection{NGC2264: Stage III}

Located in the constellation of Monoceros, the star S Mon (roughly
east of Betelgeuse) is often associated with the Cone nebula or the
Christmas-Tree nebula. It was first discovered as a cluster of young
stars, NGC2264 \citep{Herbig1954,Walker1956}, and was later found to
be a part of the giant molecular cloud complex Mon OB1
\citep[e.g.][]{Montillaud2019}. Located at a distance of
719$\pm$16\,pc \citep{Maizapellaniz2019}, it is only a little farther
away than the Orion Nebula. In the following, we show that NGC2264,
comprised of the two IR sources, IRS1 and IRS2, the Cone Nebula, and
the Spokes cluster represents the hub of an HFS that spans roughly
10\,pc is size. In Fig.\,15, we display the \herschel view of this HFS
region, at the centre of which lies the hub represented by NGC2264,
enclosing two prominent sites of intermediate-mass star formation IRS1
and IRS2 (part of Spokes Cluster). The young star S\,Mon is to the
north of this hub, and the bluish nebula below S\,Mon represents the
Fox-fur nebula. Both these objects are due to a previous event of star
formation \citep{Teixeira2008} to the one that is ongoing in the
NGC2264 HFS. A wider view in the context of Mon OB1 region is
discussed by \citet{Montillaud2019}. The main filamentary structures
F1 to F6, joining at the NGC2264 hub, are pencil-sketched in the
figure; a thorough identification of the filaments is beyond the scope
of this article. Wide-field imaging in the $^{12}$CO 3--2 and H$_2$
1-0 S(1) of this region presented by \citet{Buckle2012} shows that the
filaments joining at the hub are fertile and contain young
proto-stellar objects that are driving collimated outflows, especially
aligned along filaments F1 and F5 \citep[compare with Fig.\,3
  of][]{Buckle2012}. Even though longitudinal flows along these
filaments have not been explicitly reported in the literature so far,
the principal component analysis of the $^{12}$CO 3--2 data
\citep[Fig.\,10 of][]{Buckle2012} provides compelling evidence for
such flows. Dust continuum observations at 850\mum and 450\mum led to
the identification of NGC2264C (IRS1) as a HFS \citep{Buckle2015}, and
found that the column density along the filaments increased towards
the hub (IRS1).  These observations show large-scale ($\sim$2-5pc)
filaments, each having its own embedded population of young stars
driving collimated outflows

The hub region marked in Fig.\,15 has been the topic of numerous
studies. The IRS1 and IRS2 sources are embedded respectively in the
NGC2264-C and NGC2264-D clumps studied by \citet{Peretto2006}. A
zoom-in view of this hub region is shown in Fig.\,16. IRS1 is
well-known as the Allen's source, and is a B-type star of
$\sim$10\msun with a circumstellar disc of 0.1\msun
\citep{Grellmann2011}, and found to be a magnetically active spotted
star while IRS2 is a relatively younger Class I type object of
$\sim$150\lsun bolometric luminosity. These sources reflect the skewed
evolutionary state of near-equal mass objects mentioned several times
in this paper. Indeed IRS1 and IRS2 are the luminous sources of two
clusters which also reflect the skewness of the evolutionary
states. The cluster with a linear arrangement of young stars found
around IRS2 is called the Spokes cluster and is argued to represent
the primordial structure of the gas and dust that led to its formation
\citep{Teixeira2006}. The contours overlaid in Fig.\,16 represent the
tenth nearest neighbour densities of the young stars catalogued by
\citet{Sung2009}. They represent a rich population sampling
evolutionary states from the Class 0 to pre-main-sequence type stars
derived primarily from Spitzer observations but also from Chandra
X-ray and optical observations. We sketch purple lines in this figure
to show the large-scale filamentary structures of stellar density
joining at the hub. In this view, the linear structures identified by
\citet{Teixeira2006} indeed represent a primordial structure of the
network of finer fibres from which the Spokes cluster formed. There is
in addition to the apparent morphological evidence to argue that the
Spokes cluster represents the junction of stars that are located
inside different filaments. The radial velocities of the stars in the
Spokes cluster can be separated into at least three subgroups
\citep[see Figs.\,3 and 4 of][]{Tobin2015}, two of which are blue- and
redshifted from the component in between them. This feature also
emerges from the phase-space structure analysis of the kinematic
spectrum of the stars \citep{Gonzalez2017}. While neither of these
authors explicitly attribute the kinematic grouping to the
hub-filament systems, the data are consistent with the HFS paradigm.
The hub in NGC2264 is therefore a junction of trains of stellar
filaments. We note that these stellar filaments are offset from
observed gas filaments, the interpretations of which are not necessary
at this stage. Primordial filamentary structure influencing the
distribution of stars has also been witnessed in the DR21/W75N massive
star forming region \citep{KumarDavis2007}.

To summarise, NGC2264 represents the StageIII of our HFS paradigm,
where the following salient features can be observed: (a) A network of
fertile filaments of gas meeting at the hub, (b) a network of
filaments of stars traced by the stellar density, and the linear
arrangement of proto-stellar objects forming a junction at the Spokes
cluster, and (c) NGC2264-C and IRS2 of a relatively evolved nature
compared with NGC2264-D and IRS1, showing the skewness of evolutionary
states depicted at StageIII.

\subsection{W40 in Aquila Rift: Stage IV}

\begin{figure*}
\includegraphics[width=0.97\textwidth]{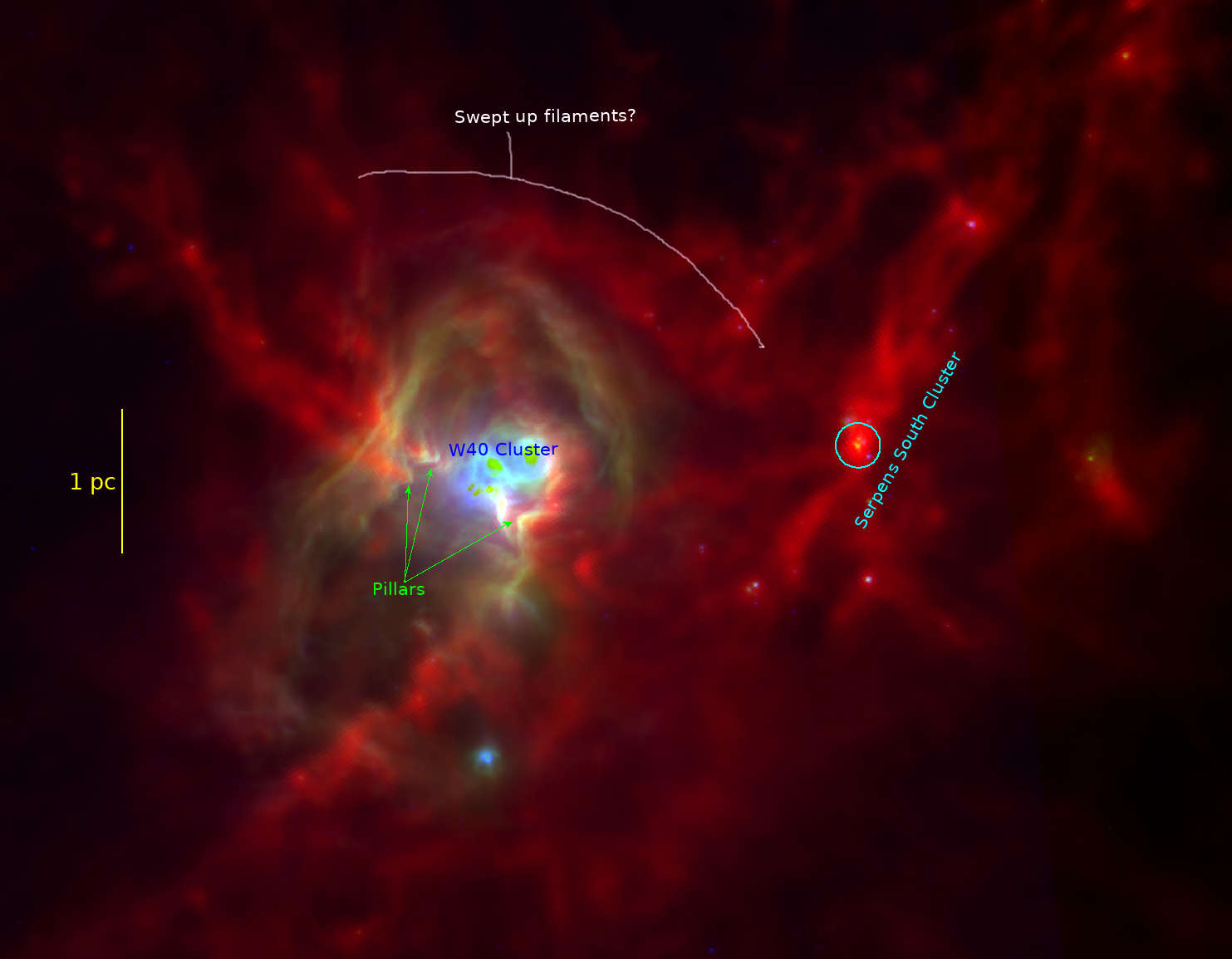}
\caption[]{W40 in the Aquila Rift as an example of StageIV: The
  formation of the W40 cluster in a hub has blown out the bipolar HII
  region \citep[seen in blue-green; see][for
    details]{Mallick2013}. The effects of radiation from OB stars
  `burning out' the hub-composing filaments can be seen as
  pillars. The ionising gas shock front sweeping up some filaments as
  it moves past it can also be seen. The colour composite uses
  \herschel SPIRE 500\mum image (tracing the filaments) as red, PACS
  70\mum as green, and {\em Spitzer} MIPS 24\mum as blue. }
\label{fig:paradigm1}
\end{figure*}

The W40 embedded cluster and HII region in the aquila rift has long
been argued to be the result of star formation in a hub at the
junction of filaments by \citet{Mallick2013}. These authors identified
the young stellar population, and showed that the YSOs are located
both in the central cluster and along the filaments. The filamentary
structure is evident both in the gas and dust and also from the YSO
density \citep[see Fig.4 of][]{Mallick2013}. However, these latter
authors argued that the star formation took place in two epochs, one
corresponding to the central cluster and the other in the
filaments. In view of the HFS scenario, the population of young stars
located in the filaments is predominantly a low-mass population that
began well before the population that quickly formed the OB stellar
cluster in the hub. This OB cluster has blown out the bipolar HII
region as shown in Fig.\,17. The scale bar of 1\,pc shown in this
figure assumes a distance of 436\,pc to the region, considered to be
the same to both W40 and Serpens South cluster \citep{Ortiz-Leon2017}.

Several features can be seen in Fig.\,17. Both W40 and Serpens south
clusters belong to the same network of filaments in the Aquila rift
region, the W40 and Serpens south representing HFS in StageIV and
StageII/III, respectively. W40 displays all the salient features of
StageIV; the OB star formation represented by the OB cluster, the
bipolar HII region created by the ionising radiation, and the effect
of burning out the natal hub-composing filaments leading to the
formation of pillar-like structures. Interestingly, the filamentary
features that are likely swept up by the expanding HII bubble may also
be represented by the 500\mum image, as marked in Fig.\,17.

The W40 and Serpens-south regions display similarities with the IRS1
and IRS2 regions within NGC2264 in terms of evolutionary differences
and physical separation. Given the lack of a fully blown HII region in
NGC2264, it likely represents an earlier evolutionary stage, though
one must consider the larger distance to NGC2264 and a denser network
of filaments.

\section{The hub-filament system compared with other models}

Models aiming to describe high-mass star formation and cluster
formation have always sought innovative ideas
\citep[e.g.][]{BBZ1998,Longmore2014,Kroupa2018} to coherently explain
observational data. Global hierarchical collapse
\citep[(GHC)][]{Vazquez2019} and conveyor belt
\citep[(CB)][]{Longmore2014} models have recently found renewed
interest, especially because of the observational support from
longitudinal flows in filaments. These models, including the HFS model
here, can derive support from several observational results. However,
the mechanisms by which the result is obtained are
different. \citet{Vazquez2019} claim that the GHC scenario is
consistent with both the competitive accretion \citep{BBZ1998} and CB
\citep{Longmore2014} scenarios, but \citet{KrumholzMcKee2020} argue
that the CB model can better explain observational data sets of ONC
and NGC6530. In contrast, these latter authors argue that the
acceleration of star formation due to large-scale collapse (GHC) or a
time-dependent increase in star formation efficiency are unable to
explain the observed data.

While a detailed comparison of the HFS paradigm with these models is
beyond the scope of this work, we highlight the major differences
in the following.

\begin{itemize}

  \item Global hierarchical collapse represents a hierarchical
    collapse of the molecular cloud where collapse occurs within
    collapses, where the most massive structure collapses at the end,
    leading to acceleration of star formation. This scenario nicely
    reproduces the sequence of star formation whereby low-mass stars
    form prior to high-mass stars and the associated acceleration when
    high-mass star formation takes place. In the HFS, we assume a
    cloud that is filamentarily in structure and that there is no
    collapse over the cloud scale, and that the sequence of star
    formation is simply due to the HFS structure. Star formation in
    hubs (especially those with large networks of filaments) may mimic
    an acceleration owing to the quick amplification of densities
    resulting from overlapping filaments. Any analysis of star
    formation rates and efficiencies will require careful
    consideration of this density amplification mechanism and
    consequent longitudinal flows.

    We envisage that the hub as an independent structure is likely to
    have more similarities with the GHC scenario. This is because the
    `two nodes of activity with an evolutionary skewness' within hubs
    is very frequently observed for which we do not have an
    explanation. These nodes are well represented by the BN/KL and
    Trapezium pair in Orion, and the IRS1 and IRS2 pair in NGC2264
    described in the previous section. Such pairs are good
    representations of the GHC scenario.

  The CB model predicts that the density of the hub remains roughly
  constant over many free-fall times, and any acceleration of star
  formation is the result of an increasing mass and not density. {
    This is in contradiction to the result shown in Fig.\,11, where
    the density increases in the hub from the pre-stellar to
    proto-stellar stages.} The HFS scenario is primarily based on the
  density amplification of hubs and therefore differs from CB in that
  respect.

  \item According to GHC, all massive clumps lead to massive star
    formation, whereas in the HFS paradigm, only those massive clumps
    that form junctions of large networks of filaments are conducive
    to the formation of massive stars. In the HFS, density
    amplification (with its associated mass increase) is responsible
    for massive star formation and not the sheer mass of the
    clump. Instead, we predict that the mass of the most massive star
    formed will be correlated with the network factor f$_{net}$ =
    $\sum_{fil=1}^{n} N_{fil}^{M_{line}} \times M_{line}^{fil}$ where
    N$_{fil}^{M_{line}}$ represent the number of filaments with a
    certain line mass M$_{line}^{fil}$. On the contrary, GHC suggests
    a correlation between the most massive star and the clump mass
    which \citet{Vazquez2019} claim to be represented by the known
    correlation of the most massive star with the cluster mass where
    it resides.

  \item Longitudinal flows within filaments are representations of
    global infall in GHC moving towards a massive clump where
    high-mass stars form. In the CB model, they serve the purpose of
    mass transport which is similar to HFS. \citet{Vazquez2019} use
    the analogy of rivers from high altitudes to lakes to describe
    longitudinal flow. On the contrary, we have used the analogy of an
    electric current driven by a potential difference. In the HFS,
    longitudinal flows are best viewed as the current in a parallel
    electric circuit, where the currents in individual circuit paths
    add up at the source which is represented by the hub. According to
    HFS, detectable longitudinal flows (end-to-end) should be absent
    in isolated individual filaments with low-mass cores. Instead,
    such filaments may have flows directed towards cores located
    within them.

  \item According to CB, the stars that form in the hub are those that
    remain as part of a bound cluster subsequent to gas dispersal. The
    HFS scenario here is different from CB in that respect. Because in
    the HFS scenario, the hub is where massive stars form, hubs evolve
    quickly, and therefore they likely result in relatively top-heavy
    mass functions. The lower mass stars and sub-stellar objects form
    slowly in the individual filaments, { and a bound cluster should
      be the result of star formation in both the filaments and the
      hub.} Observations of bound clusters clearly suggest a
    Salpeter-type mass function supporting the HFS scenario.
  
 \item In the scenario of \citet{Kroupa2018}, bursty star cluster
  formation is a result of incoming flows that are halted and
  restarted by an ionising feedback. This model also explain the
  origin of ejected O-stars. While these flows can be compared to
  longitudinal flows, the impact of feedback is unlikely to affect the
  inflow once a filamentary configuration (with a low volume filling
  factor) is considered. Formation of massive stars in a density
  amplified hub implies an enhanced cross section for dynamical
  encounters and therefore ejections, similar to that discussed by
  \citet{Kroupa2018} and \citet{oh2015}

  The initial cluster formed in the hub may provide the necessary
  gravitational potential to dynamically influence the lower mass
  stars formed in the filaments in binding them to the cluster. It is
  also possible that the stars born in the filaments will move towards
  the hub because of the angular momentum imparted to the filaments at
  StageII of HFS. This is probably the most favoured scenario
  considering the swirling spiral-arm-like structure of HFS observed
  in MonR2 \citep{TrevinoMorales2019}.
      
 \end{itemize} 

\section{Discussion}

\subsection{A hierarchy of hub-filament systems} 

In this work we searched for hubs rather than filaments. Catalogues of
filaments were produced using ATLASGAL \citep{Li2013a} and Hi-GAL
\citep{Schisano2020}. A few caveats are associated with the present
selection of HFS candidates, specifically the classification of HFS
based on line-of-sight junctions; therefore, not all of the selected
candidates are real junctions. A multi-wavelength comparison would be
necessary for improved accuracy, especially by combining higher
angular resolution data at shorter wavelengths contrasting the
\herschel data used here. A non-negligible number of sources are
located at distances beyond 8\,kpc and up to 24\,kpc. Given the
uncertainties arising from the near--far distance confusion, sources
with large distances should be viewed with extreme care, as they would
result in very long filaments of between 50 and 150pc. Such filaments
are not uncommon; indeed the giant molecular filaments or Galactic
spines such as `Nessie' are similar in length or even longer
\citep{Goodman2014}. The filaments constituting the hub sample display
a peak at lengths between 10 and 20\,pc; this is typical of giant
molecular filaments \citep{Li2013a, Zucker2018}. In the case of
Galactic spines such as Nessie, the aspect ratios are very high
(150-800). In the data analysed here, if we assumed one pixel as the
width of the unresolved filaments, the lengths in pixels imply aspect
ratios of 100-2000. Even though some of those high aspect ratios may
be real, the uncertainties in the target distance combined with the
coarse angular resolution (18\arcsec) prompts extreme caution in
viewing these aspect ratios.

Most nearby molecular clouds such as Taurus, Ophiucus, and Orion A and
B are also elongated. If these clouds were to be placed far away at
3-5\,kpc and viewed with \herschel, they would appear as giant
filaments with lower aspect ratios \citep[e.g. see Fig.\,14
  of][]{Zucker2018}, especially the Orion A and B clouds. The
Trapezium cluster can then be viewed as the hub at the junction of
Orion A and B clouds. In that sense, the filamentary structures, and
consequently the HFSs found here, can represent (a) junctions of real
giant molecular filaments with high aspect ratios, or (b) elongated
clouds of low aspect ratio.  When viewed at higher angular resolution
\citep[e.g.][]{Mattern2018}, the giant filaments and elongated clouds
are likely to resolve themselves out into networks of further finer
structures such as filaments with constant widths
\citep{Arzoumanian2011,Palmeirim2013,Arzoumanian2019} and/or velocity
coherent fibres \citep{Hacar2018}. Therefore, elongated structures
(with a range of aspect ratios) of dense gas in the cold ISM trace a
hierarchy from the Milky Way bones/Galactic spines to elongated giant
molecular clouds such as Orion, to individual filaments and fibres
within them. It appears that junctions or HFSs can form at any level
in this hierarchical distribution, even though one may naturally
expect a lower frequency of HFSs at the bones and spines level.

How is this hierarchy in filaments and hubs related to the hierarchy
of clouds, clumps, and cores? Observations show that most of the dense
gas \citep{Arzoumanian2019} within a molecular cloud is found in
filaments of high aspect ratios. Giant molecular clouds such as Orion
A and B are known to be elongated even from the earliest observations
tracing the molecular gas in them \citep{Bally1987}. Elongated
structures of dense gas such as filaments and fibres are yet to be
placed in the context of the roundish molecular clouds and
cores. However, the role of cylindrical and spherical geometries as
distinct components composing the hierarchy is quite evident when
viewing well-known targets such as Orion in the HFS scenario.

\subsection{Massive discs and pre-stellar cores}

Until now, there has been much emphasis on finding massive discs
\citep{Cesaroni2007} and massive pre-stellar cores \citep{Motte2018},
both as part of an attempt to understand the conditions that lead to
the formation of the highest mass stars such as those of 50-100\msun
or more massive objects. Both efforts have been largely unsuccessful,
especially the search for pre-stellar cores. Subsequent to the arrival
of ALMA, which provides the capacity to probe regions $<$1000\,AU,
searches for massive discs have resulted in an increased number of
detections of Keplerian- \citep{Sanchez-Monge2013,Johnston2015} or
sub-Keplerian\citep{Sanna2019}-type discs, but these are found around
stars which are at best estimated to have a mass of 20\msun. Even
though the term `massive stars' has been elusively employed in the
literature, considering 8\msun (set by free-fall time exceeding
Kelvin-Helmholtz contraction time) or 20\msun (Eddington ratio equal
to one), the real quest is to understand the formation of the
very-high-mass stars observationally catalogued in the Milky Way. In
the HFS scenario for massive star formation discussed in Sect. 5, the
above failures are perhaps expected. In the HFS scenario, the seeds of
high-mass-star formation at the level of cores or proto-stellar
fragments are not necessarily high in mass. Instead, they can be lower
mass objects located inside a density-amplified hub, where both the
accretion rates and the mass reservoir (in the HFS as a whole) are
very high, as suggested by recent observations with ALMA
\citep{Chen2019,TrevinoMorales2019}. This is, in general,
counterintuitive and at odds with the expectations of searches made so
far, where either a pre-stellar core with very high mass or a disc
around an object with very high luminosity is searched for. Contrary
to the failure of disc searches in the (sub)mm \citep{Cesaroni2017},
discs are found with a much higher frequency in infrared searches
\citep{Ilee2013} that employ CO-bandhead emission, finding Keplerian
discs with much higher masses (35\msun - 55\msun). This result perhaps
implies that well-developed disc structures may not form at earlier
stages of accretion in massive stars, or if they do, only the inner
regions ($\sim$100\,AU) are Keplerian and the outer regions
($\sim$1000\,AU) will be sub-Keplerian, as advocated by some numerical
models \citep[e.g.][]{Kuiper2011}. Indeed, a Keplerian disc is not
favourable for mass accretion without the removal of angular momentum
(see Sec.\,8.3), and therefore near-spherical accretion onto a group
of stars (as suggested by Eric Keto) fed by longitudinal flows is
perhaps the mechanism by which the most massive stars form.

\subsection{Formation of the most massive ($\ge$100\msun) stars}

The challenge in assembling $\ge$100\msun stars is to not only in the
need for a high disc accretion rate ($\sim$10$^{-3}$\msun yr$^{-1}$),
but also an envelope accretion rate and a sufficiently large massive
reservoir, allowing the disc and envelope to be replenished as they
feed the central star. Such a reservoir should be intact roughly until
most of the stellar mass is assembled, withstanding the strong
ionising radiation of the star. If the disc and envelope are
themselves not replenished, helium ash will accumulate, and the star
will move away to its death. In our scenario proposed in Sect. 5, the
filaments provide a secondary reservoir, and the flattened hub acts as
a large efficient disc where angular momentum is not an issue. The low
volume filling factors of both filaments and hub allow large masses of
dense gas to be channelled to the star in an efficient
way. Subsequently, the hub will have an even more important role in
the formation of the most massive stars, that is, in reducing
fragmentation effects. Radiation feedback and magnetic fields are both
thought to be crucial factors in controlling fragmentation as
explained in; \citet[][see Sect.2.1]{Krumholz2015}. Magnetic fields
inhibit fragmentation by inhibiting the formation of very thin and
dense accretion discs, and so disc fragmentation does not
occur. Instead, the disc slowly transports angular momentum outwards
from the centre of the star via magnetic braking. This effect works
smoothly over a larger physical scale, in contrast to radiative
heating that has its influence only within $\sim$1000\,AU of the
star. A flattened hub must be less dense than an accretion disc, but
denser than the star forming clump, which when threaded by a magnetic
field of average strength will provide stability against fragmentation
while allowing mass transfer at high rates. This may be why flattened
toroidal structures are more commonly found than Keplerian discs in
sub-mm studies of high-mass star formation
\citep[e.g.][]{Beltran2011}. Therefore, it appears that in forming the
most massive stars, there is a three-step process in mass transfer,
from filaments to hubs to envelopes/discs; whereas low-mass stars can
form easily without the hub, for example from filament fragmentation.

A magnetically threaded hub (of size $\sim$0.5\,pc) can lead to
locally ($\sim$100\,AU) high field strengths close to the star by
simply considering frozen-in-field. For example, in the case of
NGC2264 discussed in Sect.\,6.1, the CN Zeeman observations of the hub
region of NGC2264-C (IRS1) indicate marginal support \citep{Maury2012}
at the large scale. At smaller scales, observations of IRS1 show that
the star is highly magnetic and spotted, implying a stronger field
close to the star \citep{Fossati2014}. Similarly, a strong magnetic
field ($\sim$100$\mu$G) is estimated around IRS2 \citep{Kwon2011}, in
support of the role of magnetic fields in the formation of massive
stars in the NGC2264 hub.  Fossil fields from star formation are
suggested as the main mechanism to explain strong and ordered magnetic
fields in massive stars \citep{Walder2012}. Massive stars cannot be
spun down to the values observed on the main-sequence without having
long disc lifetimes or a high magnetic field strength as argued by
\citet{Rosen2012}. As we pointed out in Sect.\,8.2, the low rate of
detection of discs at earlier evolutionary stages may imply that
magnetic fields are stronger in the central regions of the hub where
the massive star forms. Hence, our result that all massive stars form
in hubs should not be surprising.

If accretion proceeds in this three-step process and a hole is punched
out by the stellar radiation in the inner $\sim$100-1000\,AU regions,
the accretion flows close to the star and/or in the disc will
naturally become strongly ionised
\citep{Keto2002,Keto2003,KetoWood2006}. Evidence that ionised gas
flows play a significant role is suggested by the Br$\alpha$ line
observations of sources in M17 \citep{Blum2004}. How much mass a star
can gain via ionised accretion flows remains to be understood,
especially accounting for simultaneous mass loss via stellar
winds. This may be key to understanding the formation of the monstrous
stars found in the Milky Way, often as optically visible stars
enshrouded by thick envelopes.

\subsection{Feedback and triggered star formation}
Thanks to the {\em Spitzer} Space Telescope, bubbles have attracted
much attention, and the Milky Way disc was dubbed a bubbling disc by
\citet{Churchwell2006}. Bubbles have been the preferred observational
targets to examine the role of feedback and its ability to trigger
star formation \citep{Deharveng2010,Palmeirim2017,Samal2018} and
specifically massive stars \citep{Zavagno2007}. \citet{Palmeirim2017}
find that $\sim$23\% of the Hi-GAL clumps are located in the direction
of the bubbles and \citet{Deharveng2010} find that 86\% of the bubbles
contain ionised gas. These observations have also renewed interest in
models of triggered star formation \citep{Walch2015,Deharveng2010},
exploring them as a mechanism for high-mass star formation. The HFS
candidate catalogue presented here includes many clumps that are
located on the rim or edge of bubbles and also represent a
hub. Therefore, even when massive star formation takes place at the
edges of bubbles, our results suggest that it happens within a
junction of three or more filamentary features located on the bubble
surfaces. Moreover, bubbles come in a variety of sizes from a few to
10-15 pc. The driving sources can be early B-type stars, young
O-stars, or supernovae, implying a large range in the energy budget
involved. In view of the HFS scenario presented here, we caution that
not all bubbles are real bubbles formed by swept up gas. In
particular, a vast majority of dense gas found in bipolar bubbles
\citep{Samal2018} may be simply confused with the hub-composing
filaments.

Triggering as a mechanism to form massive stars or simply groups of
stars should be examined with caution, accounting carefully for the
energy budget of the driving factors and the efficiency of the
resulting star formation. Indeed, \citet{Deharveng2010} point out that
only a few large bubbles (size$\ge$ 15\,pc) are candidates for
triggered star formation, in which case the driving source is often a
supernova. Lower energy processes of `sweeping' produced by B-stars
may enhance the formation of HFS or filaments, but not essentially
`trigger' more efficient star formation. Considering the sheer volume
filling factor of filaments in a molecular cloud, much of the feedback
energy can escape through inter-filamentary voids \citep[Sect. 5:
  StageIII, IV;][]{Dale2011}, while the dust-processed radiation and
ionised gas pressure helps to form and sweep the HII shells
\citep{Lopez2014} without having much effect on triggering further
star formation. Indeed, \citet{Rosen2014} find that none of the known
energy loss channels can successfully explain more than a small
fraction of the energy injected by massive stars in stellar
clusters. This implies that most of that energy is lost through
inter-filamentary voids. Swept up gas in bubbles and shells enhances
the formation of HFSs, which in turn allow massive star formation to
take place.

\subsection{Caveats and future work}

The main caveat of this observational exercise is the relatively poor
resolution of the data used to trace the HFSs. The data is sensitive
only to the most massive and nearby filaments. Next, the identified
candidates are limited by the extent of the Hi-GAL release made by the
clumps catalogue of \citet{Elia2017}, and the limitations in the
distance estimates. We foresee that a catalogue of clumps with wider
coverage and better distance estimates will be available soon that
will lead to better HFS identification.  Identification of filaments
can ideally be done in two bands with different resolution, especially
employing the PACS 160\mum band that would allow the superior spatial
resolution to be exploited in order to better resolve the HFS
structure.

While our paradigm is the result of an attempt to coherently place
several observational facts into context, the observational results
used to justify the paradigm were not made to test the
paradigm. Building upon the vast amount of high-quality data produced
by large-scale surveys, further multi-wavelength studies of both the
stars and gas simultaneously are required to test the HFS
paradigm. Such observations should be planned to carefully distinguish
features that would emerge in three main scenarios of cluster
formation, namely (a) the HFS paradigm presented here, b) GHC, and c)
the CB model. As we argue in Sect.\,7, even though all three scenarios
attempt to explain the known properties of young clusters and giant
molecular clouds, one can expect to observe specific differences
between the models. One of these is the magnitude and importance of
longitudinal flows in isolated filaments versus HFSs. The role of the
magnetic field support in the filaments and hubs, especially at
smaller physical scales, is also a crucial factor to investigate.

\section{Summary and Conclusions}

We searched for candidate HFSs in the Milky Way inner disc by
searching for filamentary structures around 35000 clumps in the Hi-GAL
catalogue detected at 3$\sigma$ and above in at least four
bands. DisPerSE software was used to accomplish this on 10\arcmin
$\times$ 10\arcmin cut-outs of 250\mum images centred on each target
clump. A hub is defined as a junction of three or more filaments on
the clump, where each filament has a minimum length of 55\arcsec (3
$\times$ FWHM of the 250\mum beam), at least 18\arcsec (one beam) of
which residing within the FWHM of the clump.

\begin{itemize}

\item Of the 34,575 clumps examined, 3703 ($\sim$11\%) are HFS
  candidates, of which $\sim$2150 (60\%) are pre-stellar and
  $\sim$1400 (40\%) are proto-stellar. We find that 144 clumps are
  saturated in one or more of the \herschel bands, all of which are
  proto-stellar in nature and are among the most active sites of star
  formation.

\item There are 26,135 non-hub clumps, of which 10,380 are located at
  the junction of two filamentary skeletons and 15,755 form the tip of
  a single filament. We find that 4,736 clumps are not associated with
  any filaments.

\item Filaments in the HFS sample are represented by mean lengths of
  $\sim$10-20pc, masses of $\sim$5$\times$10$^4$\msun, and line masses
  (M/L) of $\sim$2$\times$10$^3$\msun\,pc$^{-1}$. Filaments found
  around non-hub clumps have mean lengths of $\sim$8\,pc and masses of
  $\sim$10$^4$\msun.

\item Circularly averaged radial density profiles of all hub and
  non-hub clumps show that the column density of the hubs is enhanced
  by a factor of approximately two in pre-stellar and average
  proto-stellar sources, shooting up to a factor of about ten in the
  saturated proto-stellar sources.

\item All clumps with a luminosity L$\ge$10$^5$\lsun located within
  5\,kpc and L$\ge$10$^4$\lsun located within 2\,kpc are HFSs. Clumps
  at distances $\ge$10\,kpc are generally classified as non-hubs due
  to insufficient angular resolution to resolve the structure. This
  shows that all massive-star formation takes place in HFSs.

\end{itemize}

We propose a {\em filaments to clusters} paradigm for star formation
based on the results above.

\begin{itemize}

\item Flow-driven filaments driven by intra-cloud velocity dispersion
  or by external factors collide to form hubs. Because the junction
  happens with an offset to the centre of mass, the hub gains a small
  twist acquiring an initial angular momentum. The hub will have a
  flattened geometry, more likely an oblate spheroid with two
  gravitational centres.

\item Gravitational collapse of the hub may be delayed considering the
  relaxation time necessary for an object formed via collision of
  filaments. The hub begins collapsing at one of the two gravitational
  centres, followed by the collapse of the remaining centre. The
  enhanced density and mass of the hub provide conditions to form
  massive stars.

\item Hubs can trigger and drive longitudinal flows within filaments,
  the mass flow from which will not only bring additional mass to the
  hub, but can also replenish the reservoir in the hub as star
  formation proceeds. The radiation, ionised gas, and stellar wind
  pressures are beamed out to inter-filamentary voids by punching
  holes in the flattened hub, minimising the effects of feedback.

\item A magnetically threaded hub offers stability against
  fragmentation, favouring the formation of a group of OB
  stars. Because hubs are centrally located in HFSs, this effect
  naturally results in mass segregation. The ionised gas escaping on
  either face of the flattened hub results in bipolar HII regions,
  slowly destroying the filaments. Filament tips burnt up by radiation
  and ionisation fronts can cause the pillars of creation.

\item The net result of star formation in a HFS is a mass-segregated
  young stellar cluster, where the low-mass stars form slowly over
  10$^6$yr in the individual filaments, starting even before the hub
  formation, and high-mass stars form in the hub quickly in
  10$^5$yr. The resulting mass function is a sum of filament and hub
  (top heavy) mass functions.

\item A hub that becomes gravitationally unstable before the filaments
  can produce a cluster of low-mass stars will preferentially lead to
  the formation of isolated OB associations.
 
\end{itemize}

We compared the proposed HFS paradigm with observational
studies in the literature of two nearby well-known regions of star
formation, namely NGC2264 and W40.

\begin{itemize}

\item NGC2264 represents StageIII of the paradigm. We show that
  previously unpublished \herschel SPIRE data from the archive display
  a spectacular network of filaments feeding the NGC2264 region that
  represents a hub. The hub is composed of IRS1 (Allen source) and the
  Spokes cluster around IRS2. These two represent the two nodes within
  an oblate spheroidal hub where IRS2 is younger than the IRS1 region
  and displays the evolutionary skewness of star formation in hubs
  represented by StageIII.

\item We produce tenth nearest neighbour young stellar density maps
  using membership data from the literature. The NN maps also show
  filamentary structures merging at the hubs, and therefore the YSOs
  have not had the time to erase the structure inherited from its
  natal filaments. This result is further supported at the hub level
  as represented by the Spokes cluster that was previously claimed to
  represent primordial structure.

\item W40 displays all the characteristic features of StageIV of the
  HFS paradigm. The W40 cluster has formed in a hub that is at the
  junction of three to four main filaments. The hub is known to have
  OB stellar content that has driven a bipolar HII region that appears
  to have swept up other lower density filaments in the northern
  boundary. The main filaments display pillar-like tips as the
  radiation from the W40 cluster has destroyed the molecular
  material. We assembled a spectacular wide-field colour composite
  image to highlight these salient features.
  
\end{itemize}

\begin{acknowledgements}
  MSNK acknowledges the support from FCT - Fundação para a Ciência e a
  Tecnologia through Investigador contracts and exploratory project
  (IF/00956/2015/CP1273/CT0002). DA and PP acknowledge support from
  FCT/MCTES through Portuguese national funds (PIDDAC) by grant
  UID/FIS/04434/2019. PP receives support from fellowship
  SFRH/BPD/110176/2015 funded by FCT (Portugal) and POPH/FSE
  (EC). Herschel Hi-GAL data processing, map production, and source
  catalog generation is the result of a multi-year effort that was
  initially funded thanks to Contracts I/038/080/0 and I/029/12/0 from
  ASI, Agenzia Spaziale Italiana.  Herschel is an ESA space
  observatory with science instruments provided by European-led
  Principal Investigator consortia and with important participation
  from NASA. This work is based on observations obtained with
  Herschel-PACS and Herschel-SPIRE photometers. PACS has been
  developed by a consortium of institutes led by MPE (Germany) and
  including UVIE (Austria); KU Leuven, CSL, IMEC (Belgium); CEA, LAM
  (France); MPIA (Germany); INAF-IFSI/OAA/OAP/OAT, LENS, SISSA
  (Italy); IAC (Spain).  This development has been supported by the
  funding agencies BMVIT (Austria), ESA-PRODEX (Belgium), CEA/CNES
  (France), DLR (Germany), ASI/INAF (Italy),and CICYT/MCYT
  (Spain). SPIRE has been developed by a consortium of institutes led
  by CardiffUniv. (UK) and including: Univ. Lethbridge (Canada);NAOC
  (China); CEA, LAM (France); IFSI, Univ. Padua (Italy); IAC
  (Spain);Stockholm Observatory (Sweden); Imperial College London,
  RAL, UCL-MSSL, UKATC, Univ. Sussex (UK); and Caltech, JPL, NHSC,
  Univ. Colorado(USA). This development has been supported by national
  funding agencies: CSA(Canada); NAOC (China); CEA, CNES, CNRS
  (France); ASI (Italy); MCINN(Spain); SNSB (Sweden); STFC, UKSA (UK);
  and NASA (USA).
\end{acknowledgements}

\bibliographystyle{aa}
\bibliography{38232corr}

\end{document}